\documentclass[namedreferences]{solarphysics}

\usepackage[hyperref,optionalrh]{spr-sola-addons} 
\usepackage{graphicx}        

\newcommand{\etal}{{\it et al.}}
\newcommand{\ie}{{\it i.e.}}
\newcommand{\eg}{{\it e.g.}}

\renewcommand{\vec}[1]{{\mathbfit #1}}

\begin{document}

\begin{article}
\tracingmacros=2
\begin{opening}

\title{Energy release in driven twisted coronal loops}

\author{M.R.~\surname{Bareford}$^{1}$\sep
        M.~\surname{Gordovskyy}$^{2}$\sep
        P.K.~\surname{Browning}$^{2}$\sep
	A.W.~\surname{Hood}$^{1}$
       }

   \institute{$^{1}$ School of Mathematics and Statistics, University of St Andrews, St Andrews, Fife KY16 9SS, UK
                     email: \url{awh@st-andrews.ac.uk}\\ 
              $^{2}$ School of Physics and Astronomy, University of Manchester, Manchester M13 9PL, UK              \\
             }

\runningauthor{Bareford \etal}
\runningtitle{Energy release in twisted loops}

\begin{abstract}
In the present study we investigate magnetic reconnection in twisted magnetic fluxtubes with different initial configurations. In all considered cases, energy release is triggered by the ideal kink instability, which is itself the result of applying footpoint rotation to an initially potential field. The main goal of this work is to establish the influence of the field topology and various thermodynamic effects on the energy release process. Specifically, we investigate convergence of the magnetic field at the loop footpoints, atmospheric stratification, as well as thermal conduction. In all cases, the application of vortical driving at the footpoints of an initally potential field leads to an internal kink instability. With the exception of the curved loop with high footpoint convergence, the global geometry of the loop change little during the simulation. Footpoint convergence, curvature and atmospheric structure clearly influences the rapidity with which a loop achieves instability as well as the size of the subsequent energy release. Footpoint convergence has a stabilising influence and thus the loop requires more energy for instability, which means that the subsequent relaxation has a larger heating effect. Large-scale curvature has the opposite result: less energy is needed for instability and so the amount of energy released from the field is reduced. Introducing a stratified atmosphere gives rise to decaying wave phenomena during the driving phase, and also results in a loop that is less stable.
\end{abstract}
\end{opening}
\tracingmacros=0

\section{Introduction}
\label{sec_intro}

The solar corona is thought to be heated to temperatures of millions of degrees Kelvin by dissipation of stored magnetic energy. The details of the processes of energy storage and dissipation 
remain contentious, and it is likely that the corona is heated by a combination of mechanisms \citep{pade12}. A very plausible scenario, especially for heating in Active Regions, is that the corona 
is heated by the combined effect of many small flare-like events known as "nanoflares" \citep{park88}.  Thus, understanding of flares, especially smaller events, contributes to solving the coronal 
heating problem.  It is essential to understand the heating of loops, which are the main building blocks of the coronal magnetic field \citep{real14}.

Twisted magnetic fields are ubiquitous in the solar corona, and twist is associated with free magnetic energy. New  magnetic flux ropes emerging from below the solar surface should already be twisted, 
and further twisting is produced by photospheric footpoint motions with vorticity.  In essence, the magnetic fields that permeate the solar corona acquire free energy due to the convective motions that 
take place in and around the loop footpoints: \ie, where the field intersects the photosphere. There is an increasing body of observational evidence for solar flares occurring in twisted coronal loops 
\cite[\eg][]{srie10, kure13, kuch14}, and high resolution observations from HiC show untwisting of "braided" fields associated with energy release \citep{cire13}. Here, we focus on the energy release within a single coronal 
loop containing magnetic field which is twisted by vortical photospheric motions. The primary emphasis is on modelling microflares and similar events, but the combined effect of many such events 
with different magnitudes also provides an effective coronal heating mechanism \citep{brva03,bare10,bare11}.

Previously,  numerical simulations have shown that magnetic energy release, sufficient for coronal heating above active regions, can occur as a consequence of an ideal instability 
\citep{broe08,hooe09,bote11,bare13}.  As the free magnetic energy increases, the coronal loop becomes more and more susceptible to the kink instability \citep{hood92}. This instability, although 
in itself an ideal process, triggers the formation of multiple current sheets and leads  to magnetic reconnection at many sites within the loop volume.  The  magnetic reconnection causes energy to be 
released  from the field and heats the coronal plasma.

The earlier models of this process start with a field configuration that is known to be already  unstable (\ie slightly beyond the threshold for linear ideal kink instability). In addition, the 
field is usually modelled as a simple straight cylinder where the field strength depends on the distance from the loop axis only. In reality, coronal magnetic  fields must expand from localised 
photospheric sources -  despite the much-discussed phenomenon of  ``constant cross-section'',  which now seems more likely to reflect the plasma emission \citep{klim00, pebi12}. Furthermore, coronal loops 
are also inevitably curved. 

These simplifications may affect the dynamics and energetics of twisted loops, artificially restricting or exaggerating
 the amount of magnetic energy released. As well as the field geometry, features involving atmospheric physics, which are expected to influence how well magnetic energy is thermalised within the loop volume, 
are also usually ignored. For example, most previous work  uses a simplified adiabatic energy equation, although the effects of thermal conduction in cylindrical unstable loops have been considered by 
\cite{bote11}. 

Recently, \cite{gore14} investigated the kink instability and magnetic reconnection in more realistic configurations with large-scale curvature (\ie loop-like fluxtubes) and atmospheric stratification. Although the key focus of that study is particle acceleration, it has revealed some features not present in idealised cylindrical models, such as the loop contraction, asymmetric current distribution and variation of the energy release along the loop. Hence, the loop topology and various thermodynamic effects can substantially affect the energy release process and, therefore, they need to be investigated in more detail.
 
The aim of this paper is to consider heating in twisted coronal loops, involving more realistic models, and, along the way, we explore the effects of removing various idealisations and simplifications which have been 
adopted in previous works. Does the basic mechanism of an ideal kink instability triggering the formation of  fragmented current sheets, leading to significant dissipation of magnetic energy, persist in a 
realistic loop model? If so, how are the dynamics and energetics affected by the magnetic field geometry and the plasma physics incorporated within the modelling? Therefore, we investigate  a family of four 
loop models which allow us to compare and contrast the effects of field  geometry, both loop curvature and field expansion/convergence. In contrast to previous models which assume an initial unstable equilibrium, 
the twisted fields are established by rotating the footpoints of the field. Furthermore, in the case of the more realistic curved loop models, we explore the effects of incorporating thermal conduction, and also embed the loop in a gravitationally-stratified atmosphere. In order to study the effect of stratification, we introduce an additional model, with the atmosphere stratified in both density and temperature, thus representing  a loop whose lower layers are embedded in the upper chromosphere.

The paper is structured in the following manner. Section \ref{sec_loop_configs} describes the different loop configurations used for the nonlinear magnetohydrodynamic (MHD) simulations. Section \ref{sec_num_code} presents the 
numerical code and Section \ref{sec_num_res} focuses on how results change as the loop model is made more and more realistic: \eg, how the energy release is affected by the addition of curvature, or how the results 
change when a stratified atmosphere is used. Finally, in the last section, the results are summarised and our conclusions are given.

\begin{figure}[h!]
  \center
  \includegraphics[width=0.9\textwidth,clip=]{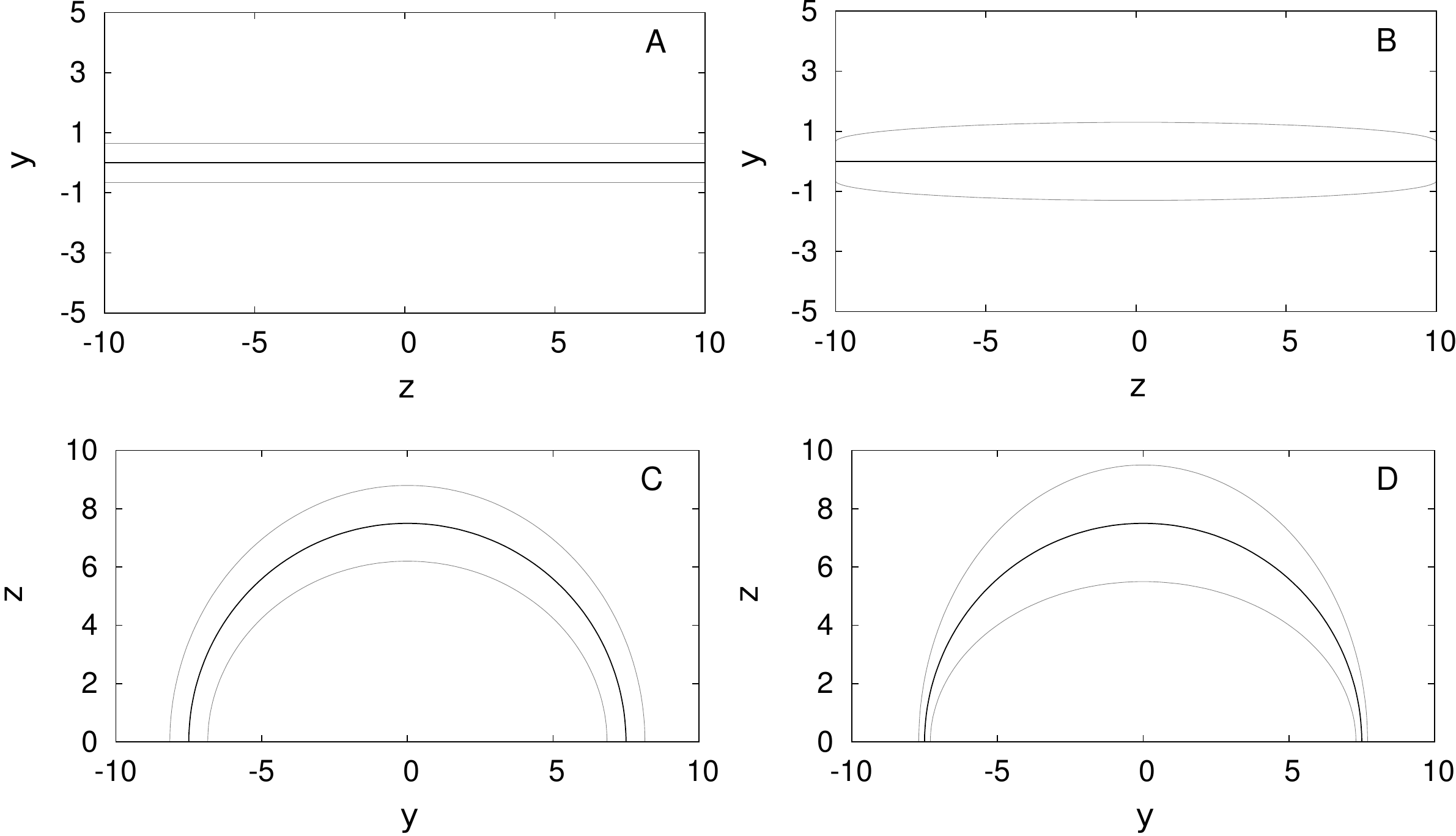}
  \caption{\small{Initial magnetic field geometry for loops A (top left), B (top right), C (bottom left) and D (bottom right).}}
  \label{fig_loops}
\end{figure}

\begin{figure}[h!]
  \center
  \includegraphics[width=0.9\textwidth,clip=]{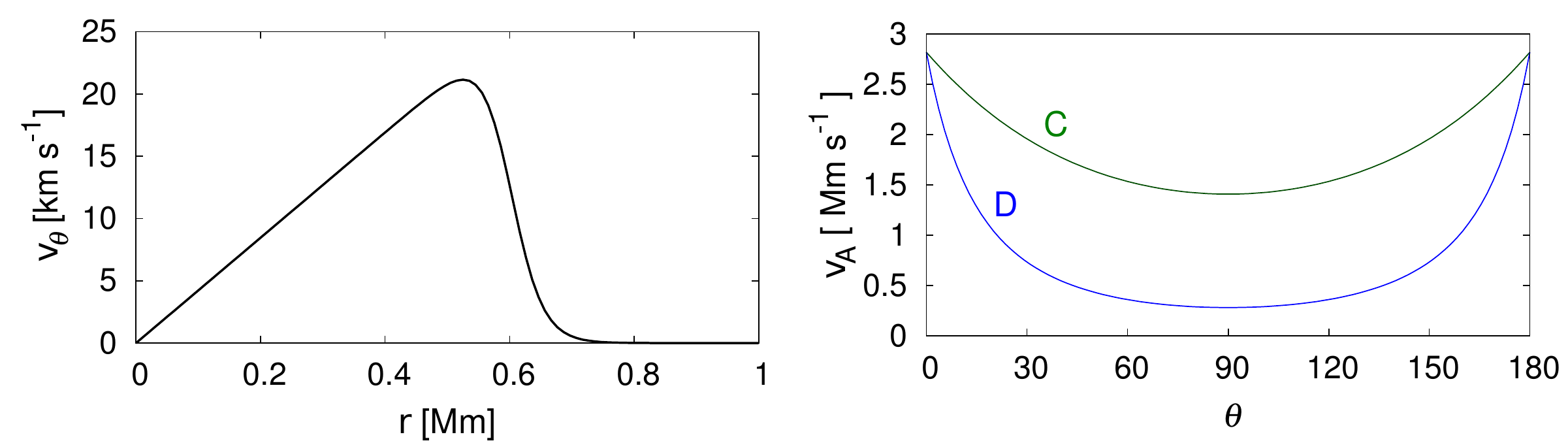}
  \caption{\small{{\bf Left}, the footpoint driving function for $0.1t_\mathrm{tw}\,{\lesssim}\,t\,{\lesssim}\,0.9t_{tw}$. {\bf Right}, the Alfv\'en speed along the central field line of loops C (green) and D (blue) --- $\theta=90$ is the position at the loop apex.}}
  \label{fig_drv_spd}
\end{figure}

\section{Loop configurations}
\label{sec_loop_configs}

The primary goal of this study is to investigate the effect of field geometry on the energy release in twisted loops.
Hence, our simulations cover four separate field configurations, see Figure~\ref{fig_loops}. In models A and B, we consider straight fluxtubes using a domain with dimensions $x=\pm 5$~Mm, $y=\pm 5$~Mm, $z=\pm 10$~Mm, with the fluxtubes initially orientated along the z-axis. The first fluxtube (A) has a constant cross section, whereas model B has a field converging towards the footpoints: the field strength at the footpoints is twice the value at the centre (or apex) of the fluxtube. The purpose of loop B then is to show the influence of field convergence with respect to the kink instability and subsequent heating. Loops C and D have more realistic configurations, featuring large-scale curvature and are orientated differently compared to the straight loops. The curved loop domains have dimensions $x=\pm 10$~Mm, $y=\pm 10$~Mm, $z=0$~Mm~-~$20$~Mm, with both footpoints residing at the $z=0$ boundary, representing the chromosphere. Loop C is constructed such that the field converges towards the footpoints in a manner comparable to loop B (both have $B_{\mathrm{ftp}}\,/B_{\mathrm{apx}}\,{=}\,2$): the differences in results between these two loops should therefore reveal the impact of curvature. The footpoint positions for loop D are the same as those for loop C, and so these two loops exhibit a similar level of curvature; however, loop D is given five times the level of footpoint convergence ($B_{\mathrm{ftp}}\,/B_{\mathrm{apx}}\,{=}\,10$): the intention here is to continue to explore the impact of field convergence, but within the context of loop curvature. 

The initial field for loop A is uniform and has a $z$-component only, while in loops B, C, and D the initial field is constructed using two point sources located outside the model domain:
\begin{eqnarray}  
  \vec{B}(\vec{r}) & = & B_s\, \Bigg(\frac{\vec{r} - \vec{s}_1}{|\vec{r} - \vec{s}_1|^3} ~-~ \frac{\vec{r} - \vec{s}_2}{|\vec{r} - \vec{s}_2|^3}\Bigg)\,,
  \label{eqn_cnv_b}  
\end{eqnarray}
where $\vec{r}$ is a position vector within the simulation volume, $\vec{s}_1$ and $\vec{s}_2$ are the positive and negative point sources; $B_s$ is a simple scaling constant and is given a specific value such that the axial field strength at the footpoints is some multiple of the field strength at the apex. In model B these point sources are located outside the $z_{\min}$ and $z_{\max}$ boundaries, while in C and D they are both located below $z_{\min}$ boundary. Table \ref{tab_convergence} gives the point source coordinates and scaling factors required to implement the converged field geometries.
\begin{center}
  \begin{table}[h!]
    \caption{The point source coordinates and scaling terms required for field convergence (see eqn. \ref{eqn_cnv_b}).}
    \label{tab_convergence}
    \begin{tabular}{ c  c  c  c  c  c }
      \hline
      \textbf{Loop} & $\vec{s}_1$ & $\vec{s}_2$ & $B_s$ & $B_{\mathrm{ftp}}$ & $B_{\mathrm{apx}}$ \\ \hline      
      \textbf{A} & n/a & n/a & n/a & 1 & 1\\ 
      \textbf{B} & (0, 0, -21.4) & (0, 0, 21.4) & 114 & 1 & 0.5\\ 
      \textbf{C} & (0, -7.5, -27.1) & (0, 7.5, -27.1) & 1480 & 1 & 0.5\\ 
      \textbf{D} & (0, -7.5, -4.2) & (0, 7.5, -4.2) & 17.9 & 1 & 0.1\\
      \hline    
    \end{tabular}
  \end{table}
\end{center}
The density and temperature are initially uniform in models A, B, C and D: $n\,{=}\,1.2\times10^{15}\,\mathrm{m}^{-3}$ and $T\,{=}\,4\times10^3\,\mathrm{K}$. In addition, model D is
considered with a gravitationally stratified atmosphere (as model D*), see Section~\ref{sec_stratified_atmosphere}.

For all loops discussed in this paper, the magnetic field is under the line-tied condition ($\eta =0$, \ie $\partial \vec{B}/\partial t ={\bf 0}$ when $\vec{v} ={\bf 0}$) at the ``foot-point'' boundaries.
Furthermore, all loop simulations begin with a potential field.
The magnetic twist necessary for instability is created by rotating the plasma located at the footpoints; rotation vortices are, for the curved loops, centered at $x\,{=}\,0$, $y\,{=}\,\pm 7.5$, $z\,{=}\,0$. \mbox{The rotational driving} varies in space and time (Figure \ref{fig_drv_spd}, left):
\begin{eqnarray}
  v_{\theta}(r,t) &=& r\,\omega_0 \, \frac{1-\tanh \left( \frac{r-r_{\mathrm{fr}}}{r_{\mathrm{fb}}} \right)}{2} \times \tanh\left(\frac{t}{t_{\mathrm{sw}}} \right) \times \frac{1-\tanh\left(\frac{t-t_{\mathrm{tw}}}{t_{\mathrm{sw}}} \right)}{2}\,,
  \label{eqn_drv_ftp}
\end{eqnarray}
where $r$ is the radial distance from the vortex centre, $\omega_0\,{=}\,0.015$ is a scaling factor, $r_{\mathrm{fr}}\,{=}\,0.6\,R_0$ is the footpoint radius, $r_{\mathrm{fb}}\,{=}\,0.05\,R_0$ is the footpoint boundary thickness, $t_{\mathrm{sw}}\,{=}\,20\,t_{\mathrm{A}}$ is the \textit{switching} time, and $t_{\mathrm{tw}}$ is the characteristic twisting time. There is a slightly different arrangement for the straight loops: the vortex centres are located at $(0,0,\pm10)$ and the two footpoints are driven in opposite directions. Note, the dimensionalising factors are $R_0\,{=}\,1\,\mathrm{Mm}$ and $t_{\mathrm{A}}\,{=}\,0.35\,\mathrm{s}$ (see Section \ref{sec_num_code} for further details). The important parameters are $\omega_0$ and $t_{\mathrm{tw}}$: the product of these two terms, multiplied by two, gives the total twist after footpoint rotation has stopped: \eg, for $t_{\mathrm{tw}}\,{=}\,750$ the twist is ${\sim}\,6\pi$. Each loop is twisted for a time sufficient to cause an instability. The Poynting flux generated by the driving adds energy to the field; if the driving is continued through the unstable phase it becomes difficult to assess the energy released as the loop relaxes to a lower energy state. Thus, $t_\mathrm{tw}$ changes according to the loop: the values of the twisting time for the five loops are 750 (A), 650 (B), 400 (C) and 600 (D).

The level of field convergence determines how the Alfv\'en speed varies with height. For loop A there is no convergence and $v_{\mathrm{A}}\,{\approx}\,2800\,\mathrm{km}\,\mathrm{s}^{-1}$ for all $z$, which is also true for the footpoints of loops B-D. Hence, the peak driving velocity ($\sim21\,\mathrm{km}\,\mathrm{s}^{-1}$) is sub-Alfv\'enic (Figure \ref{fig_drv_spd}, right) for all four loops. 

\section{Numerical code}
\label{sec_num_code}

All numerical simulations were performed using LARE3D -- 3D Lagrangian remap MHD code \citep{arbe01}. This code is based on a Lagrangian remap scheme: the Lagrangian part, which is done using a second-order accurate predictor-corrector method, deforms the grid such that it moves with the plasma. The advantage of this technique is that additional physics, such as thermal conduction and shock capturing, can easily be incorporated into the code. The remap stage involves the mapping of the plasma properties (\eg, density, velocity, magnetic field) back to the original Cartesian grid; monotonicity is preserved through the use of \citet{vanl97} gradient limiters.

LARE3D solves the following resistive MHD equations,
\begin{eqnarray}
  \label{eqn_lare_mhd_mass}
  \frac{\partial\rho}{\partial t} & = & -\nabla\cdot(\,\rho\vec{v}\,)\,,\\
  \nonumber \\ 
  \label{eqn_lare_mhd_force}
  \frac{\partial}{\partial t}\big(\,\rho\vec{v}\,\big) & = & -\nabla\cdot(\,\rho\vec{v}\vec{v}\,)\,\,+\,\,\frac{1}{\mu_0}\Big(\,\nabla\times\vec{B}\,\Big)\times\vec{B}\,\,-\,\,\nabla{P}\,\,-\,\,\rho g \vec{z}+\,\,\nabla\cdot{\vec{\sigma}}\,\,\,\,\\
  \nonumber \\ 
  \label{eqn_lare_mhd_induction}
  \frac{\partial\vec{B}}{\partial t} & = & \nabla\times\Big(\,\vec{v}\times\vec{B}\,\Big)\,\,-\,\,\nabla\times\Bigg(\eta\frac{\nabla\times\vec{B}}{\mu_0}\Bigg)\,,\\
  \nonumber \\ 
  \label{eqn_lare_mhd_energy}
  \frac{\partial}{\partial t}\big(\,\rho\epsilon\big) & = & -\nabla\cdot(\,\rho\epsilon\vec{v}\,)\,\,\,-\,\,P\,\nabla\cdot\vec{v}\,\,+\,\,\eta J^{\,2}\,\,+\,\,\nabla\cdot\vec{q}\,\,+\,\,\vec{\varepsilon}\,\vec{\sigma}\,,
\end{eqnarray}
with specific energy density,
\begin{eqnarray}
  \label{eqn_lare_mhd_epsilon}
  \epsilon & = & \frac{P}{(\gamma-1)\,\rho}\,,
\end{eqnarray}
and current density
\begin{eqnarray}
  \label{eqn_lare_mhd_j}
  \vec{J}  & = & \frac 1{\mu_0} \vec{\nabla}\times \vec{B}\,,
\end{eqnarray}
where $\rho$ is the mass density, $\vec{v}$ is the plasma velocity, $\vec{B}$ the magnetic field, $P$ the thermal pressure, $\eta$ is the resistivity, $\gamma\,{=}\,5/3$ is the ratio of specific heats, $\mu_0$ is the magnetic permeability, and $\vec{q}$ is the conductive heat flux. Radiation is ignored in this study, but it is not expected to have a significant impact, since the radiative timescale is much longer than the magnetic relaxation timescale. The gravity is ignored ($g=0$) in models with uniform atmosphere (A-D), but included in model D* with atmospheric stratification. The last terms of \mbox{equations \ref{eqn_lare_mhd_force} and \ref{eqn_lare_mhd_energy}} feature tensors, and are required to simulate shock heating; these terms are defined in \mbox{Section \ref{sec_shock_res}}.

We normalise the variables in the MHD equations using reference values suitable for a coronal active region:
\begin{eqnarray}
  \nonumber r & = & \frac{r^*}{R_0}\,,\,\,\,\,\,\,\rho\,\,=\,\,\frac{\rho^*}{\rho_0}\,,\,\,\,\,\,\,B\,\,=\,\,\frac{B^*}{B_0}\,,
\end{eqnarray}
where asterisks denote the unnormalised MHD variables, $R_0=1$~Mm, $\rho_0=2\times10^{-12}$~kg~m$^{-3}$ and $B_0=4.47\times 10^{-3}$~T. Other variables are expressed as follows;
\begin{eqnarray}
  \nonumber L & = & \frac{L^*}{R_0}\,,\,\,\,\,\,\,t\,\,=\,\,\frac{t^*}{t_{\mathrm{A}}}\,,\,\,\,\,\,\,v\,\,=\,\,\frac{v^*}{v_{\mathrm{A}}}\,,\,\,\,\,\,\,P\,\,=\,\,\frac{P^*}{P_0}\,,
\end{eqnarray}
where $v_{A}\,{=}\,B_0/\sqrt{\mu_0\rho_0}$ is the reference Alfv{\'e}n speed, $t_{\mathrm{A}}\,{=}\,R_0/v_{\mathrm{A}}$ is the reference Alfven time, and $P_0\,{=}\,B_0^{\,2}/\mu_0$ is the reference pressure. The specific energy density, current density and resistivity ($\epsilon$, $J$ and $\eta$) also have reference variables that can be expressed in terms of $R_0$, $\rho_0$ and $B_0$:
\begin{eqnarray}
  \nonumber \epsilon_0 & = & \frac{B_0^2}{\mu_0\rho_0}\,\,=\,\,v_{\mathrm{A}}^2\,,\,\,\,\,\,\,J_0\,\,=\,\,\frac{B_0}{\mu_0 R_0}\,,\,\,\,\,\,\,\eta_0\,\,=\,\,\mu_0 R_0 v_{\mathrm{A}}\,.
\end{eqnarray}
Hence, for the chosen values of $R_0$, $\rho_0$ and $B_0$, $t_{\mathrm{A}}\,{\approx}\,0.36\,\mathrm{s}$, $v_{\mathrm{A}}\,{\approx}\,2800\,\,\mathrm{km}\,\mathrm{s}^{-1}$ and 
$\eta_0\,{\approx}\,1.1\pi\,{\times}\,10^6\,\Omega\,\mathrm{m}$.

In all four cases, the simulation features two stages \citep[see also][]{gore13}: twisting until the fluxtube becomes kink-unstable and magnetic relaxation after the kink instability.
The first stage is done using ideal MHD: Ohmic dissipation and conduction are absent from the energy equation. The footpoint driving eventually induces a kink instability that initiates a release of magnetic energy. Just before this point, the simulations are restarted with the resistive and conductive terms now included in the MHD equations. Any heating will reduce density and thereby increase the local Alfv\'en speed, which then leads to a decrease in the time step under the CFL condition. For the converged loops the magnetic field is strongest at the footpoints, and so the driving will inevitably create the highest currents at these locations too. The burst of Ohmic heating caused by the switch on of resistive MHD can be so intense as to cause cavitation at the $z$ boundaries; the result is that the time step becomes too small for the simulation to make progress. Fortunately, this issue can be avoided through the use of thermal conduction (Section \ref{sec_therm_con}), assuming that the plasma below the footpoints is maintained at a constant cold temperature equivalent to $4000\,\mathrm{K}$.

Throughout the loop volume there is a background resistivity of $\eta_{\mathrm{b}}\,{=}\,3 \times 10^{-7}$, except when the current reaches or exceeds a threshold ($j_{\mathrm{crit}}\,{=}\,2$), at which point an anomalous resistivity, $\eta_{\mathrm{c}}\,{=}\,0.002$, is applied. The lower the value of $j_{\mathrm{crit}}$ the greater the percentage of the loop interior that will be assigned an anomalous resistivity when the simulation is restarted in resistive mode. We chose a current threshold of two, since this results in $10\%$ of loop A contributing to Ohmic heating immediately after the restart.

The computational domain is a 3D staggered grid: physical variables are not calculated at the same place for each cell in the domain, which improves numerical stability and allows conservation laws to be included in the computation. There are some differences between the straight and curved loops as regards grid dimensions and limits. The straight loop simulations are run at a grid resolution of $128^2\,{\times}\,256$, whereas for the curved loops it is $256^2\,{\times}\,512$. The differences in grid volume and resolution between the straight and curved loop simulations mean that, along the $x$ and $y$ axes, the straight loops are better resolved by a factor of two; however, the curved loops have double the resolution along the $z$ axis.

\subsection{Thermal conduction}
\label{sec_therm_con}
The conductive heat flux (equation \ref{eqn_lare_mhd_energy}) is implemented using Braginskii parallel conduction \citep{brag65} which becomes isotropic where the magnetic field is weak: 
\begin{eqnarray}
  \label{eqn_lare_mhd_heat_flux}
  \vec{q} & = & \Bigg(\kappa\,\frac{\vec{B}}{|\vec{B}|^2\,+\,b_{\mathrm{min}}^{\,2}}\cdot\nabla T\Bigg)\vec{B} ~+~ \kappa\,\frac{b_{\mathrm{min}}^{\,2}}{|\vec{B}|^2\,+\,b_{\mathrm{min}}^{\,2}}\, \nabla T,
\end{eqnarray}
where $\kappa\,{=}\,\kappa_0\,T^{5/2}\,\,\big(\kappa_0\,{=}\,10^{-11}\big)$ and $b_{\mathrm{min}}\,{=}\,0.001$. Conduction results in the transfer of internal energy across the grid and is calculated implicitly over half the time step, using successive over relaxation. Before equation \ref{eqn_lare_mhd_heat_flux} can be applied the code variables must be converted from normalised units to SI units. This is done by amending the kappa constant,
\begin{eqnarray}
  \label{eqn_amended_k0}
  \kappa_0 & = & \frac{B_0^{\,4}}{L_0\,\rho_0^3}\,\frac{1}{\mu_0^2}\,\Bigg(\frac{m_p}{k_B}\Bigg)^{7/2}\,10^{-11}\,.
\end{eqnarray}
Then the changes in internal energy must be converted back into normalised units before the code can proceed.

\subsection{Shock resolution}
\label{sec_shock_res}

LARE3D uses shock viscosity \citep{wilk80} to capture the heating effect of shocks, see the last term of the energy equation (\ref{eqn_lare_mhd_energy}). It is expressed as the product of the rate of strain tensor,
\begin{eqnarray}
  \label{eqn_lare_strain_tensor}
  \varepsilon_{ij} & = & \frac{1}{2}\Bigg(\frac{\partial\,v_i}{\partial\,x_j}\,\,+\,\,\frac{\partial\,v_j}{\partial\,x_i}\Bigg)\,,
\end{eqnarray}
and the shock tensor,
\begin{eqnarray}
  \label{eqn_lare_shock_viscosity_tensor}
  \sigma_{ij} & = & \rho\,l_{\mathrm{s}}\,\Big(\nu_1\,c_{\mathrm{ms}}\,\,+\,\,\nu_2\,l_{\mathrm{s}}\,|s|\Big)\,\Bigg(\varepsilon_{ij}\,\,-\,\,\frac{1}{3}\,\delta_{ij}\,\nabla\cdot\vec{v}\Bigg)\,,
\end{eqnarray}
where $c_{\mathrm{ms}}\,{=}\,\sqrt{c_{\mathrm{s}}^2 + v_{\mathrm{A}}^2}$ is the magnetosonic speed, $l_{\mathrm{s}}$ is the distance across a grid cell in the direction normal to the shock front, $s$ is a similarly localised strain rate (the subscripts $i$ and $j$ denote the different spatial coordinates), and the shock viscosity coefficients $\nu_1\,{=}\,0.1$ and $\nu_2\,{=}\,0.5$ are constants  (real viscosity is set to zero). 

The method of using shock viscosity to represent heating due to shocks is known to yield physically valid results. Recently, \citet{bare14} conducted a detailed investigation of shock handling within LARE3D. They have shown that, for kink-unstable loop simulations, LARE3D viscous heating is consistent with Petschek reconnection and slow-mode shocks. Unlike the anomalous resistivity, controlled by the externally-defined critical current, the dissipation due to shock viscosity is defined internally, by the velocity field in the simulation domain. 

\section{Results of numerical simulations}
\label{sec_num_res}
This section presents the results taken from the loop simulations (A-D). Some properties are given in SI units; these are length, temperature and density. Velocities are expressed in either $\mathrm{km}\,\mathrm{s}^{-1}$ or $\mathrm{Mm}\,\mathrm{s}^{-1}$. Times are given in units of the Alfv{\'e}n time, whereas energies and currents are given in normalised units.

\begin{figure}[h!] 
  \center 
  \includegraphics[width=0.9\textwidth,clip=]{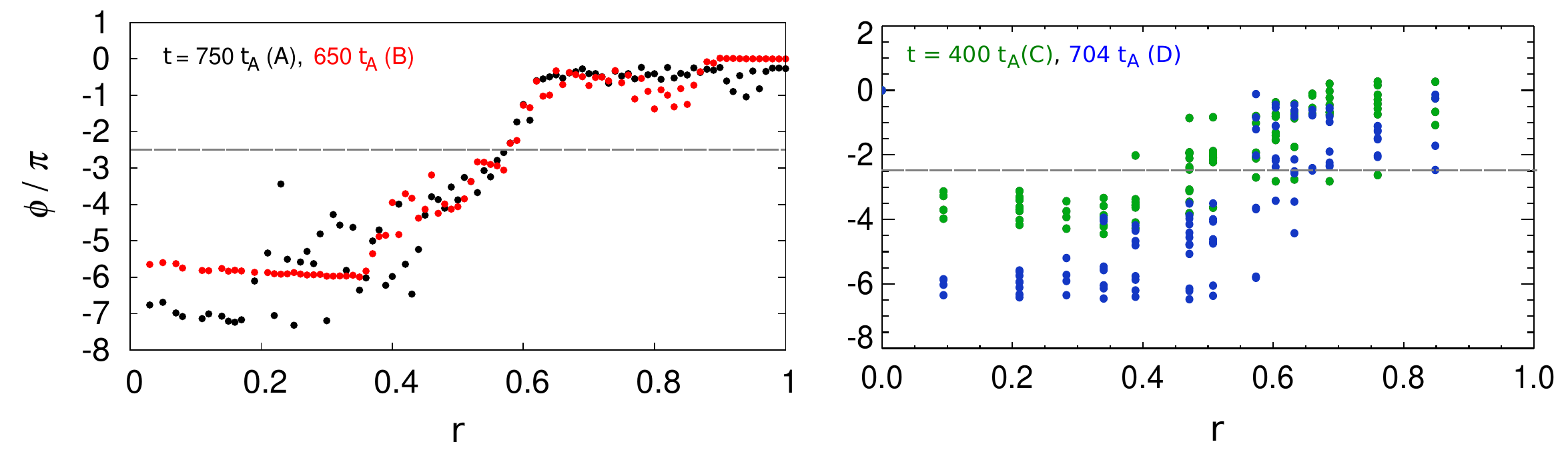}
  \caption{\small{\textbf{Left}, the magnetic twist in units of $\pi$ as a function of radial distance from footpoint centre for the straight loops, A (black) and B (red). \textbf{Right}, the same plots but for the curved loops, C (green) and D (blue). The twist values were determined numerically just before instability onset, see time labels. The twist threshold for an ideal kink instability ($\phi\,{\approx}\,2.49\pi$) as calculated by \mbox{\citet{hopr79}} is given by the dashed horizontal line.}}
  \label{fig_twist}
\end{figure}
\begin{figure}[h!]
  \center
  \includegraphics[width=0.9\textwidth,clip=]{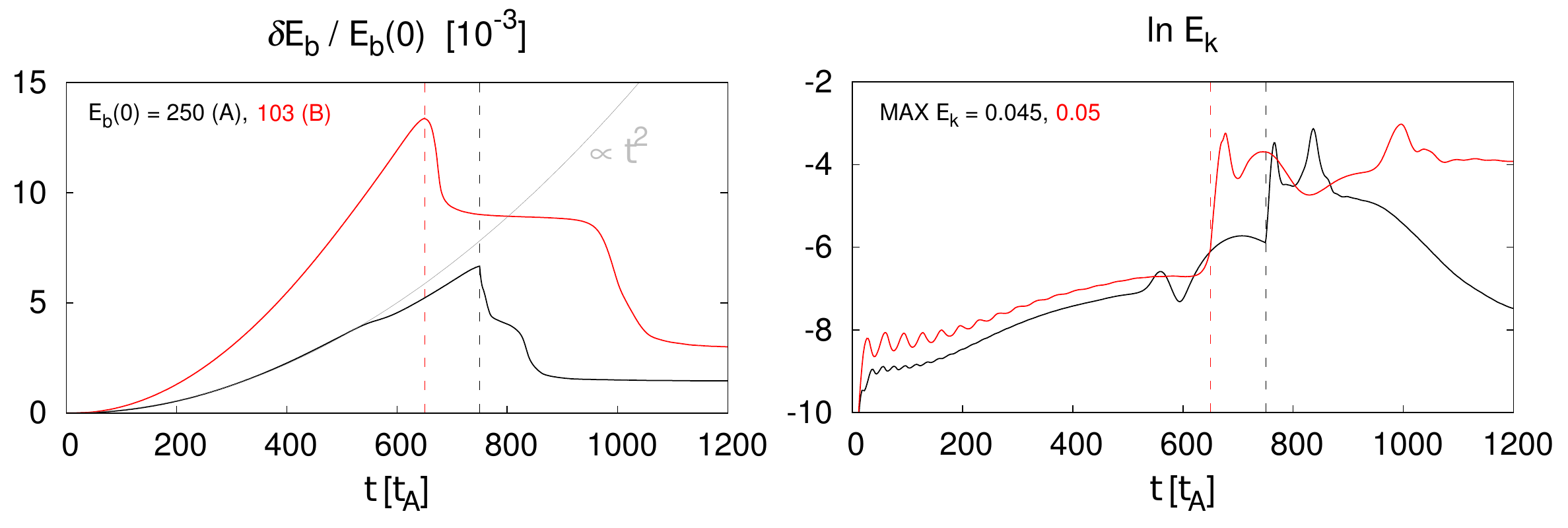}
  \caption{\small{The change in magnetic energy normalised by the initial value, $E_{\mathrm{b}}(0)$ (left) and the natural logarithm of the kinetic energy (right) for loop A (black) and for loop B (red). Resistive MHD was switched on at $t\,{=}\,750\,t_{\mathrm{A}}$ for loop A and $t\,{=}\,650\,t_{\mathrm{A}}$ for loop B , \ie, just before instability. Energies are volume intergrated.}}
  \label{fig_en_db_straight}
\end{figure}
\begin{figure}[h!]
  \center
  \includegraphics[width=0.9\textwidth,clip=]{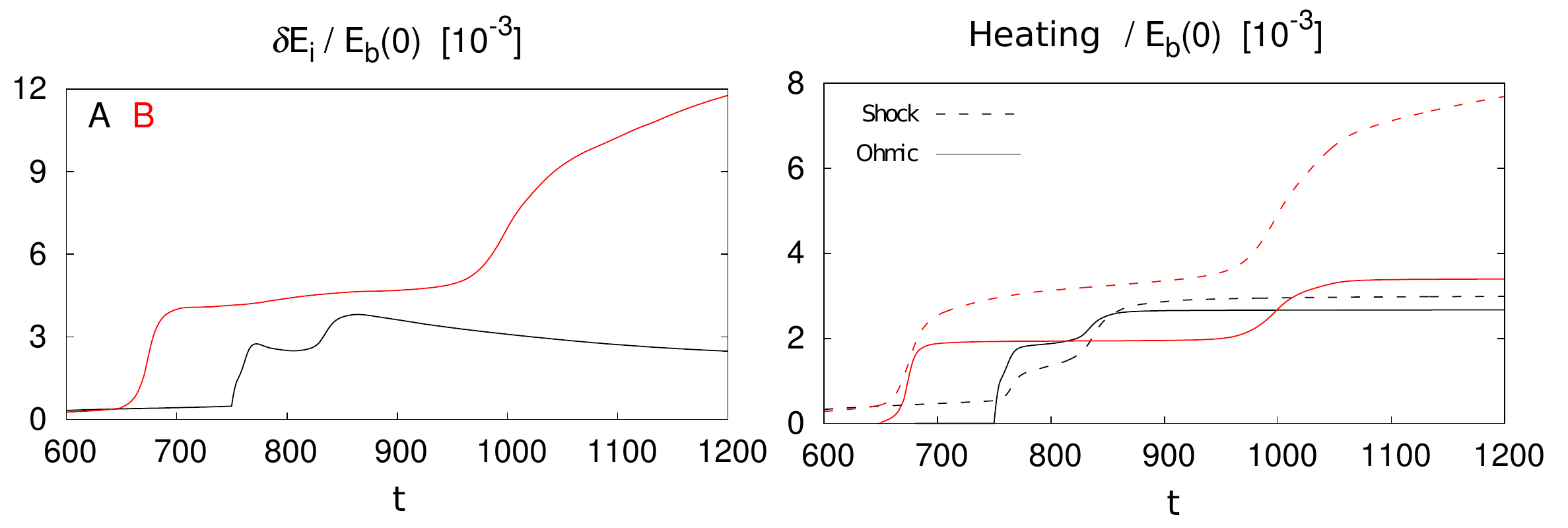}  
  \caption{\small{The cumulative Ohmic and shock heating (left) and the change in internal energy (right) for loop A (black) and for loop B (red). Energies are volume intergrated and normalised by the initial magnetic energy, $E_{\mathrm{b}}(0)$.}}
  \label{fig_en_di_straight}
\end{figure}
\begin{figure}[h!]
  \center
  \includegraphics[width=0.9\textwidth,clip=]{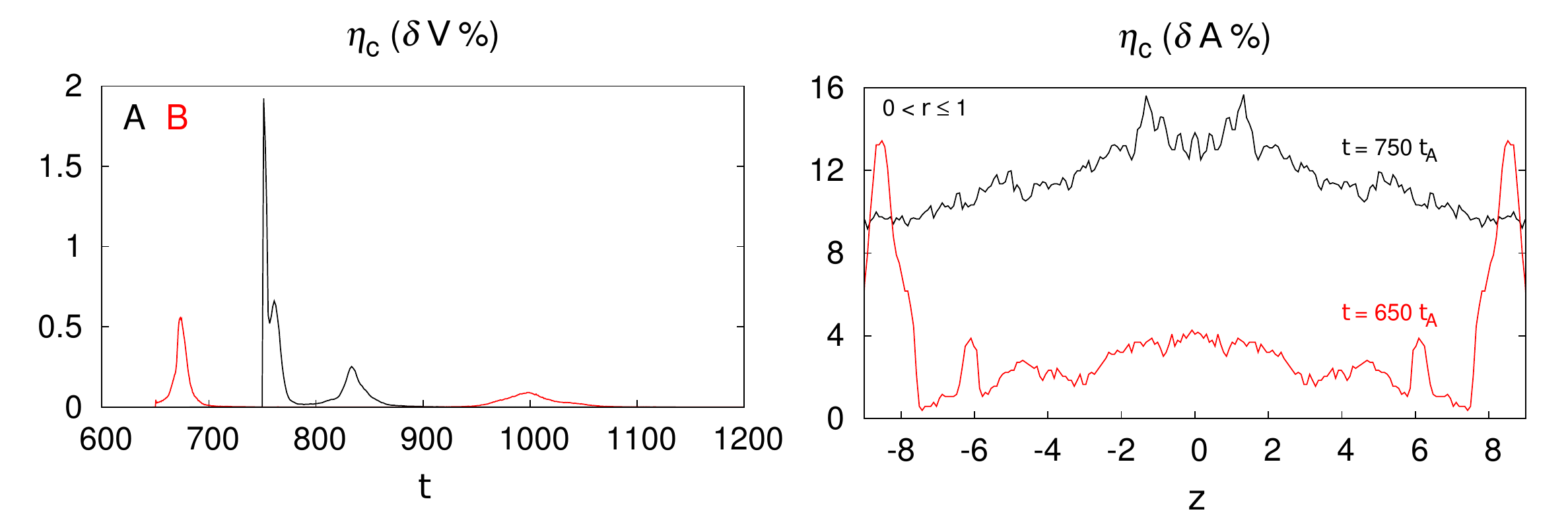}
  \caption{\small{\textbf{Left}, the percentage of the grid assigned anomalous resistivity for loop A (black) and for loop B (red). \textbf{Right}, the percentage of the cross sectional area ($0\,{<}\,r\,{\le}\,1$) assigned anomalous resistivity at $t\,{=}\,750\,t_{\mathrm{A}}$.}}
  \label{fig_eta_crit_straight}
\end{figure}
\begin{figure}[h!]
  \center
  \includegraphics[width=0.9\textwidth,clip=]{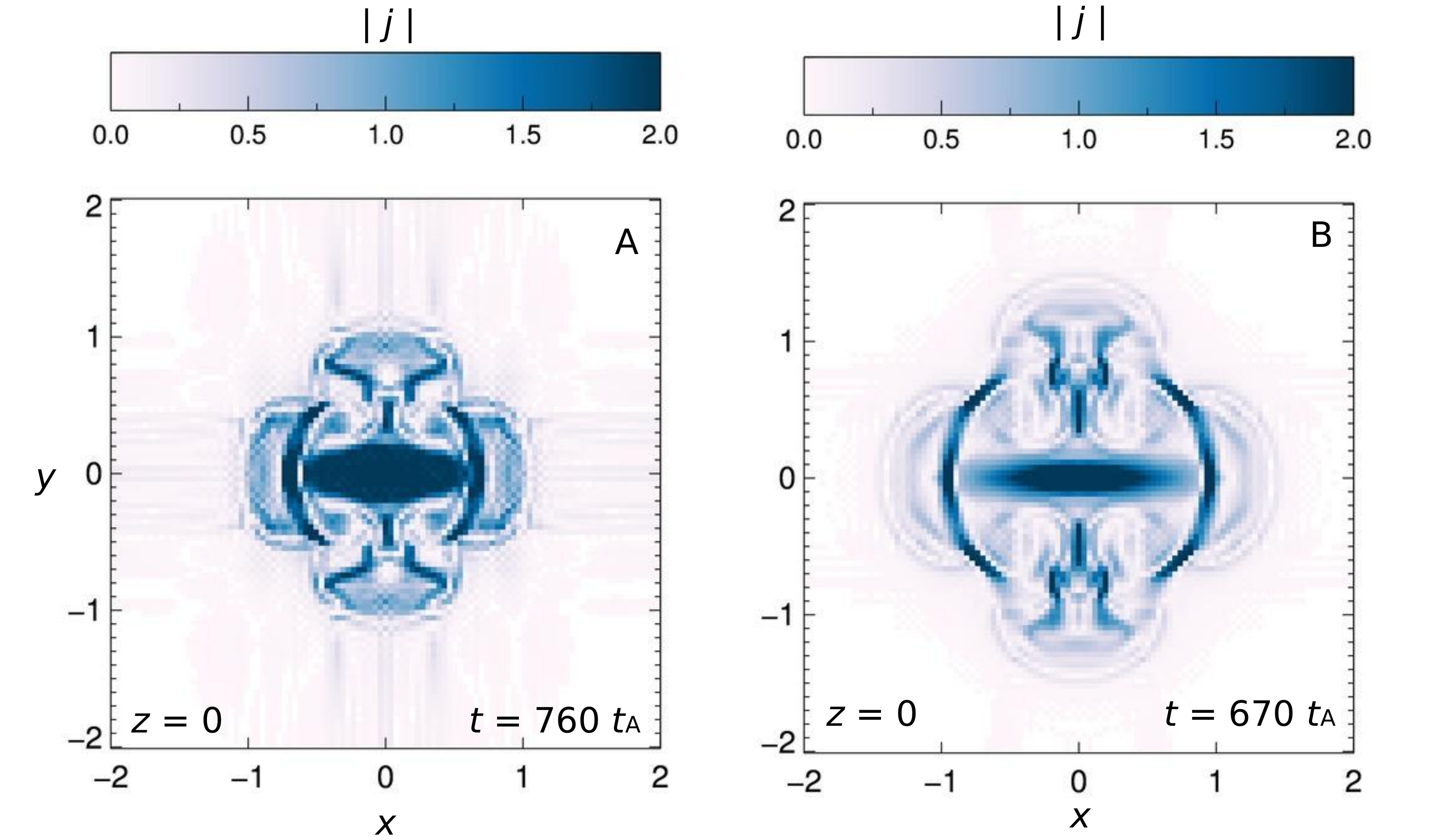}
  \caption{\small{The current density magnitude (dark regions correspond to the highest currents) over the $x$-$y$ plane at $t\,{=}\,751\,t_{\mathrm{A}}$ and $z\,{=}\,2.4$ (loop A, left) and $z\,{=}\,0.6$ (loop B, right).}}
  \label{fig_jmag_straight}
\end{figure}
\begin{figure}[h!]
  \center  
  \includegraphics[width=0.9\textwidth,clip=]{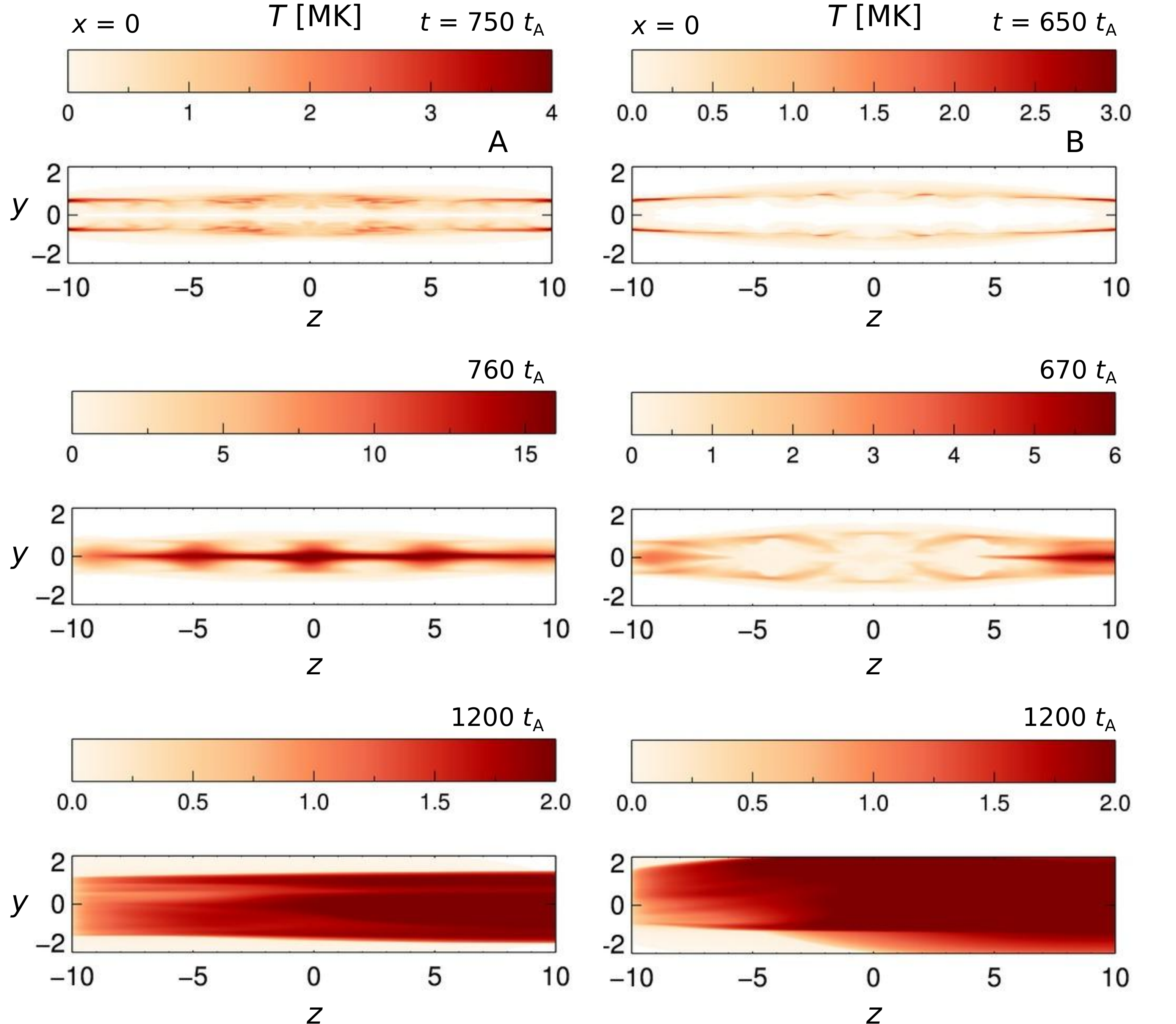}
  \caption{\small{The temperature over the $y$-$z$ plane ($x\,{=}\,0$) at times $t\,{=}\,750$ (top), $760$ (middle) and $1200\,t_{\mathrm{A}}$ (bottom) for loop A (left) and for loop B (right).}}
  \label{fig_temp_a_b}
\end{figure}
\begin{figure}[h!]
  \center
  \includegraphics[width=0.9\textwidth,clip=]{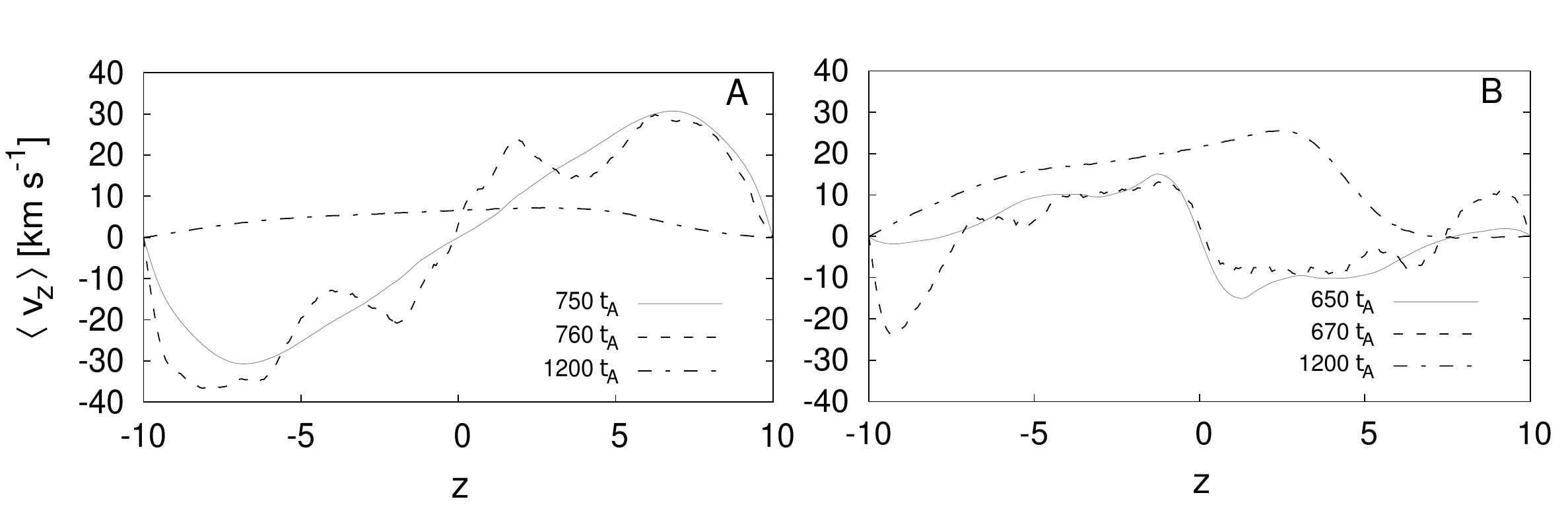}
  \caption{\small{The average velocity parallel to the $z$-axis at three times for loops A (left) and B (right). The velocity component $v_z$ is averaged over $0\,{<}\,r\,{\le}\,0.8$ for each $z$ coordinate.}}
  \label{fig_vz_straight}
\end{figure}

\subsection{Kink instability}
\label{sec_kink_instability}
First, we check that the driving phase has generated sufficient twist consistent with the ideal kink instability. Figure \ref{fig_twist} is created by following field lines from specific points on the postive footpoint. The starting points are distinguished by the distance from the footpoint centre, located at (0,0,-10) for the straight loops and (0,-7.5,0) for the curved; and the magnetic twist ($LB_{\theta} / rB_z$) is calculated numerically by keeping track of how many times a field line wraps around the initial loop axis before it reaches the negative footpoint.
The twist measurements were taken at a time close to instability onset. It can be clearly seen that the level of twist for both straight loops exceeds $2.49\pi$, the result obtained by \citet{hopr79} which is applicable to a straight loop of aspect ratio ten ($\,{=}\,L/R$): hence, this value should only be treated as a necessary condition for instability. At radii less than $0.4$, the magnetic twist is approximately 5-7$\pi$: this is consistent with the average twist on the instability threshold (where the twist is single-signed) reported by \citet{bare11}. However, the curved loops are significantly less twisted: in particular, the twist profile for loop D is never greater than $3\pi$ and is only above the Hood and Priest threshold when $0.35\,{\lesssim}\,r\,{\lesssim}\,0.4$. These results suggest that curvature exerts a destabilising influence: \ie, the average absolute twist at instability onset is lower than it is for the straight loops.

\subsection{Footpoint convergence}
\label{sec_convergence}
In order to investigate the importance of footpoint convergence we compare the results for loops A and B. The former begins with a uniform straight field of $|\,\vec{B}\,|\,{=}\,1$, whereas the latter begins with a field that has less energy, since $|\,\vec{B}\,|$ only rises to one at the footpoints. 

Figures \ref{fig_en_db_straight} and \ref{fig_en_di_straight} present the changes in volume-integrated energies and, with the exception of the kinetic energy logarithm, these terms have been normalised by the initial magnetic energy, $E_{\mathrm{b}}(0)$, in order to allow a meaningful comparison between the two straight loops. The vortical footpoint driving \mbox{(equation \ref{eqn_drv_ftp})} leads to instability for both cases; however, the converged straight loop (B) is the first to achieve instability. For this reason, resistive MHD is switched on earlier, see the vertical dash lines in Figure \ref{fig_en_db_straight}.

Proportionally, the driving phase adds more azimuthal field to loop B, even though it is driven for less time: furthermore, after $t\,{=}\,650\,t_{\mathrm{A}}$, loop B undergoes two bursts of energy release. The first occurs, as expected, at the end of the driving phase, then the loop stabilises for $300\,t_{\mathrm{A}}$ before undergoing a slightly greater burst of energy release (similarly, loop A shows a quasi-stable period centred on $800\,t_{\mathrm{A}}$). By the end of the simulation, loop B has released more than double the magnetic energy released by loop A. As a result, the converged loop produces higher levels of kinetic and internal energy. Incidentally, the energy released by loop B increases only marginally (${\sim}\,8\%$) should the driving phase be extended to $t\,{=}\,750\,t_{\mathrm{A}}$ (as is the case with loop A): the instability is delayed until this later time after which the magnetic energy declines to a relaxed state, with no intervening stable period.

It is expected that the driving should generate a linear increase in azimuthal field over time: hence, the increase in magnetic energy should show a $t^2$ dependence. This relationship is confirmed for loop A by the dashed line in Figure \ref{fig_en_db_straight} (left), although, there is clearly a faster increase in magnetic energy when the field converges at the footpoints. After the driving phase, the natural logarithm of the kinetic energy rises again when resistive MHD is switched on (the gradient of $\ln E_{\mathrm{kin}}$ is twice that of $\gamma$, the growth rate of the instability, since the velocity of the perturbed plasma is proportional to $e^{\gamma t}$). The wave-like form present from the start of the simulations is a consequence of the fact that the driving speed is supersonic; however, by $t\,{=}\,400\,t_{\mathrm{A}}$, these oscillations have diffused away, which indicates that the field is now evolving through a sequence of equilibria \citep{mele05}. Overall, the numerical dissipation, measured as a percentage of the total initial energy, is $0.11\%$ for both straight loops.

The levels of Ohmic and shock heating are roughly equal for the unconverged loop (A), but when the field is converged the shock heating becomes steadily greater as the simulation progresses: in fact, Ohmic heating only increases during the two periods of magnetic energy decline. 

Figure \ref{fig_eta_crit_straight} shows the percentage of the grid assigned anomalous resistivity (left), which, of course, is zero until the start of the resistive phase. The plot on the right gives the location of supercritical currents as a function of $z$: the percentage value refers to the portion of a circle ($r\,{=}\,1$) whose origin follows the $z$-axis that has $\eta\,{=}\,0.002$ (Figure \ref{fig_drv_spd}, left).

Inevitably, the driving creates high currents at the footpoints --- these are not shown in Figure \ref{fig_eta_crit_straight} (right) in order to reveal how $\eta$ changes around the loop apex. Nevertheless, for loop A, $90\%$ of the loop volume assigned anomalous resistivity occurs within $-9\,{\le}\,{z}\,{\le}\,9$, whereas loop B has $82\%$ of the anomalous resistivity within this range.

We now investigate the current structures existing shortly after instability onset (Figures~\ref{fig_jmag_straight} and \ref{fig_jcd}). Similar to previous simulations of kink-unstable loops \citep{broe08,hooe09,bote11,bare13}, the current sheets in the cylindrical models A and B form a helical ribbon structure around the kinking fluxtubes. In addition, there is a strong currect along the fluxtube axis ($x=y=0$). This current structure has elliptic cross-section with its main axis rotating along the $z$-axis; the total rotation angle just after the kink is approximately the same as the total magnetic twist angle. In the model A (with initially cylindrical 
fluxtube) the current density is neary uniform along $z$-axis: it is slightly higher in the central region of the fluxtube (\ie around $z=0$), while in the fluxtube with converging field (model B) it also has high current densities near foot-points \cite[see also][]{gobr11,gobr12}.

Next, we examine how heat is distributed within the two straight loops at three times, $t\,{=}\,750\,t_{\mathrm{A}}$ (instability onset), $760\,t_{\mathrm{A}}$ (immediately after the instability) and $1200\,t_{\mathrm{A}}$ (when the field has relaxed to a lower energy state), see Figure \ref{fig_temp_a_b}. Leading up to the instability (top row), hot plasma forms shells around the fluxtubes, apparently, corresponding to the current density concentrations. At this stage the temperatures are also comparable. However, immediately after the instability, temperatures for loop A increase by a factor of four, with the highest temperatures forming along the axis. The radial spread of heating at certain locations along the $z$-axis reveals where the loop has kinked. Contrastingly, loop B shows the strongest heating near the footpoints, as well as in the shell around the fluxtube; the temperatures are roughly double those before the instability. As the loops relax, thermal conduction acts to smooth out the temperatures, yielding an average value of $2\,\mathrm{MK}$.

Finally in this section, we look at how $v_z$ varies along the loop axis (Figure \ref{fig_vz_straight}). The velocity is averaged over the loop cross-section ($0\,{<}\,r\,{\le}\,0.8$) for the same times as the temperature plots. Before the instability (solid line), the stongest axial flows for loop A are away from the apex. These flows diminish as the instability progresses, resulting in a residual left to right flow by the time loop A has relaxed. Loop B has a more complicated flow pattern, consistent with heating sources located at $z\,{=}\,\pm\,7.5$. Over time the flow becomes chaotic, but the end result is similar to loop A, albeit with a higher rightward flow.

\subsection{Loop curvature}
\label{sec_curvature}

\begin{figure}[h!]
  \center
  \includegraphics[width=0.9\textwidth,clip=]{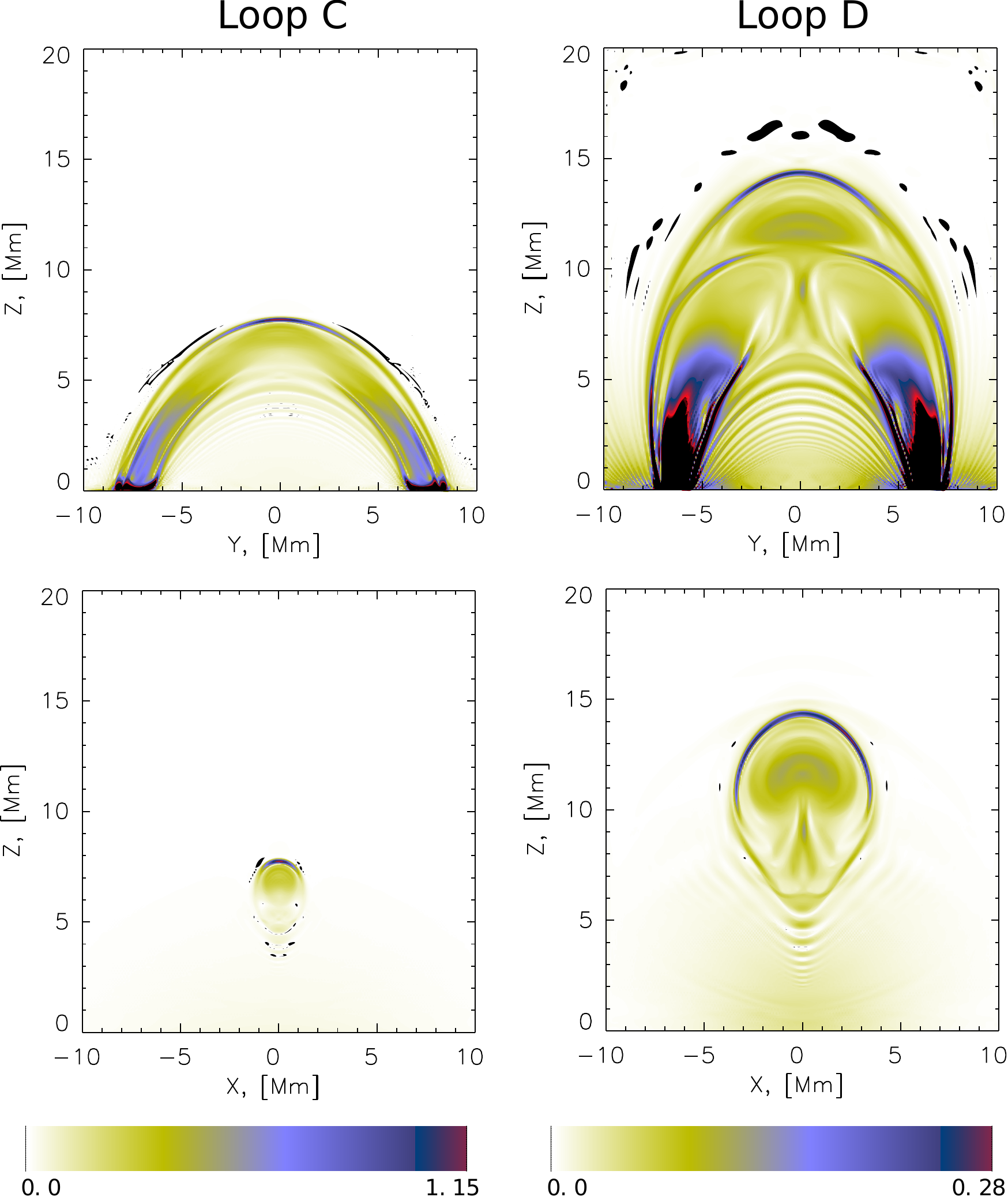}
  \caption{\small{Current density distributions just before the kink instability in models C and D. Upper panels show loop mid-planes ($x=0$), lower panels show central cross-sections ($y=0$). Current density scales are 
shown in units of $j_0=3.6\times 10^{-3}$~A~m$^{-2}$.}}
  \label{fig_jcd}
\end{figure}

The obvious difference between curved loops (models C and D) and cylindrical fluxtubes is the geometry of current density distribution. In the straight loop (model A) currents are nearly uniformly distributed along the loop, while
in the converging fluxtube (model B) the current density is, as one would expect, higher near the footpoints. However, in both models the structure has high degree of cylindrical symmetry. At the same time, in curved loops 
(models C and D) the current distributions do not have cylindrical symmetry. Thus, just before the instability occurs, the current is concentrated in a thin shield above the loop top (see Figure~\ref{fig_jcd}). In addition, 
there are current concentrations close to footpoints due to the field convergence.

\begin{figure}[h!]
  \center
  \includegraphics[width=0.9\textwidth,clip=]{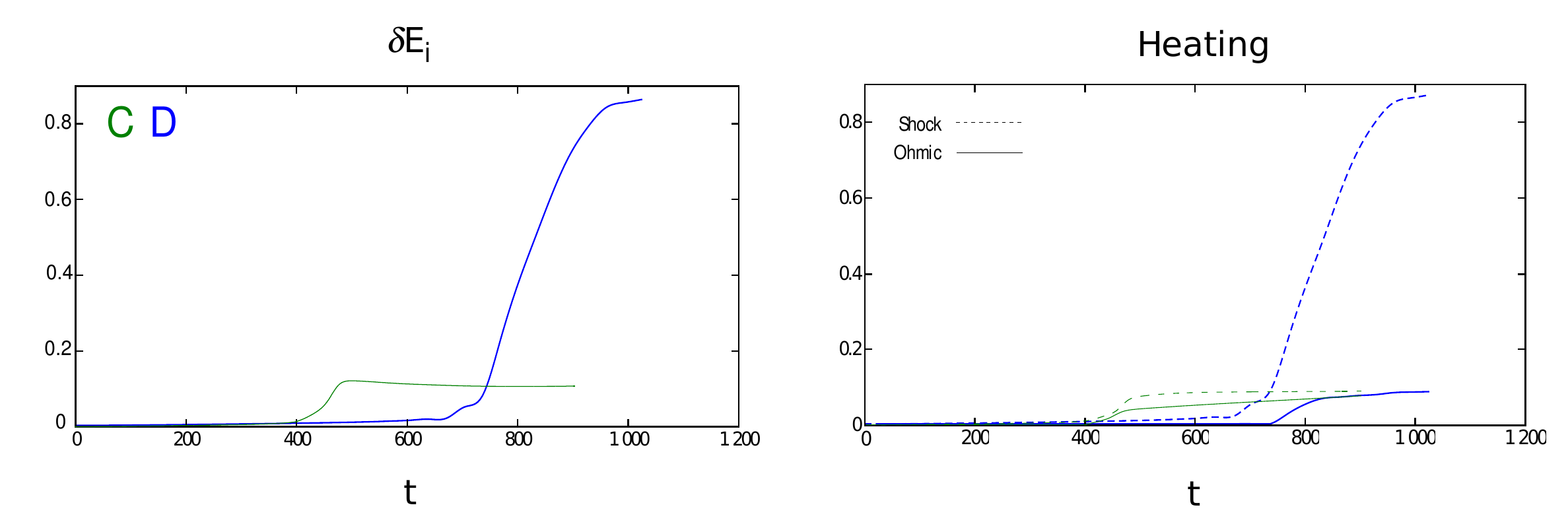}    
  \caption{\small{The cumulative Ohmic and shock heating (left) and the change in internal energy (right) for loop C (green) and for loop D (blue). Energies are volume intergrated.}}
  \label{fig_en_di_curved}
\end{figure}

\begin{figure}[h!]
  \center
  \includegraphics[width=0.9\textwidth,clip=]{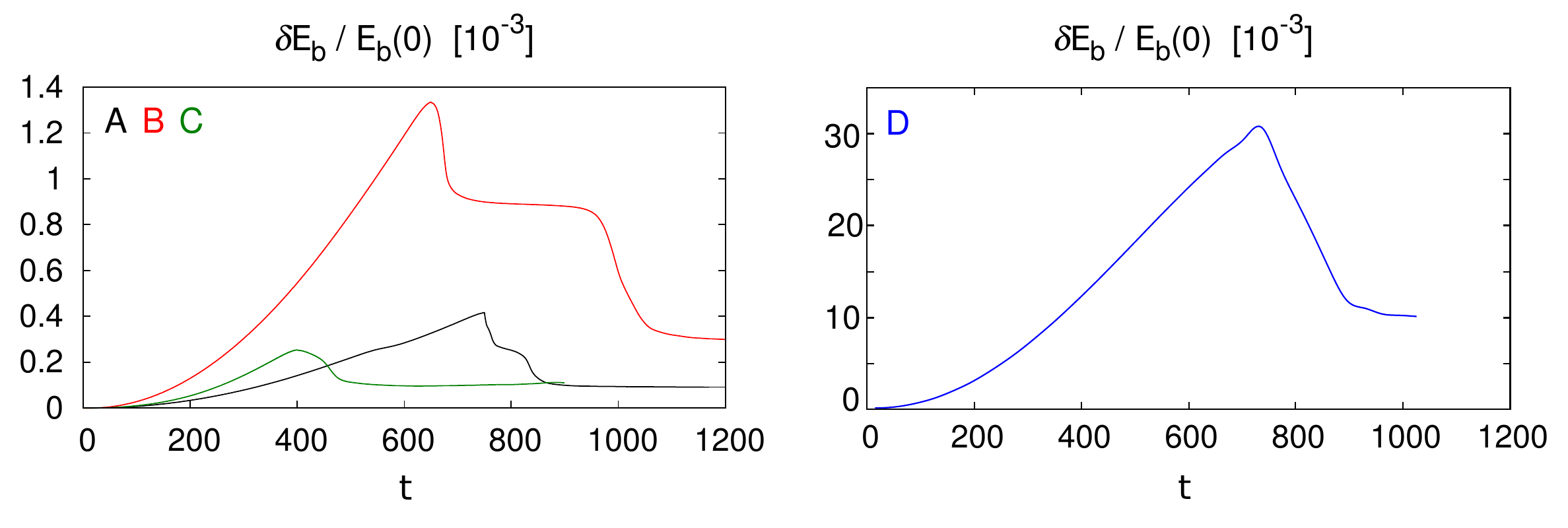}  
  \caption{\small{{\bf Left}, the volume-integrated change in magnetic energy normalised by the initial value, $E_{\mathrm{b}}(0)$ for loops A (black), B (red) and C (green). {\bf Right}, the same plot but with the change in magnetic energy for loop D (blue). The weightings used for loops A and B also take into account the difference in grid volume between the straight and curved loop simulations: the latter being sixteen times the former.}}
  \label{fig_en_db}
\end{figure}

Loop D has the most highly converged footpoint field, which means the field strength at the apex is the weakest (all loops have the same footpoint field strength). The result of this is that loop D has by far the smallest volume-integrated field energy --- over ten times smaller than the total field energy for the other curved loop (C). Hence, the energy plots for these two models are strikingly different. Figures \ref{fig_en_db_curved} and \ref{fig_en_di_curved} are not normalised by the initial integrated field energy, since other plots would appear almost flat by comparison (Figure \ref{fig_en_db}, right).

As with the straight loops, increasing the convergence, increases the rate at which the loop accrues free magnetic energy during the driving phase; but, within the context of curvature, greater convergence appears to delay the instability by some $200\,t_{\mathrm{A}}$. Loop D has to be driven for at least $700\,t_{\mathrm{A}}$ before an instability will occur, whereas loop C encounters an instability at $400\,t_{\mathrm{A}}$. 

In addition, the two curved loop simulations appear to differ as regards the nature of the instability. Loop C is consistent with the kink-like instabilities seen for the straight loops: there is a swift drop in magnetic energy coincident with rises in heating (both Ohmic and shock) and internal energy. Furthermore, the unstable phase is also accompanied by peaks in kinetic energy. Loop D on the other hand, shows a more gradual drop in magnetic energy (Figure \ref{fig_en_di_curved}), which does not correspond to the rise in Ohmic heating; instead the similarly gradual increase in internal energy is caused by shock heating. Surprisingly, the kinetic energy remains high even as loop D settles into a lower energy state. Driving loop C for the same amount of time as loop D (\ie, $t_{\mathrm{tw}}\,\approx\,600\,t_{\mathrm{A}}$) does little to alter how the magnetic energy changes during the simulation. The instability occurs at roughly the same time as before ($400\,t_{\mathrm{A}}$), but the energy released is halved: the continued driving interferes with how the instability plays out.

To understand the impact of curvature, we compare the normalised change in magnetic energy from the straight loops with loop C (Figure \ref{fig_en_db}, left). Comparing the red and green plots, we see that curvature has throttled the rate at which driving adds magnetic energy to the loop. At the same time however, large-scale curvature has made the loop more susceptible to instability: a loss of magnetic energy occurs much earlier. 

In the right panel of Figure \ref{fig_en_db}, we add the result from loop D to show how continuing to increase footpoint convergence almost completely cancels the effect of curvature. The fall in magnetic energy still happens earlier
than for the straight loops, but the growth in magnetic energy before instability far outstrips that seen in the other simulations. The normalised increase in magnetic energy is eight times higher than that for loop B.

We now turn our attention to showing whether or not the results from loop D are indeed consistent with a kink instability. Although, Figure \ref{fig_twist} (right) shows that loop D is sufficiently twisted (stronger than the loop C), we also see that the changes in magnetic energy and the kinetic energy logarithm (Figure \ref{fig_en_db_curved}, blue lines) are different from the forms seen for other kink-unstable loops. Specifically, the decline in magnetic energy is noticeably slower and not accompanied by a rapid change in kinetic energy. It seems that increasing the convergence (compared to loop C) has prolonged the transition to a low-energy state.

\begin{figure}
  \center
  \includegraphics[width=0.9\textwidth,clip=]{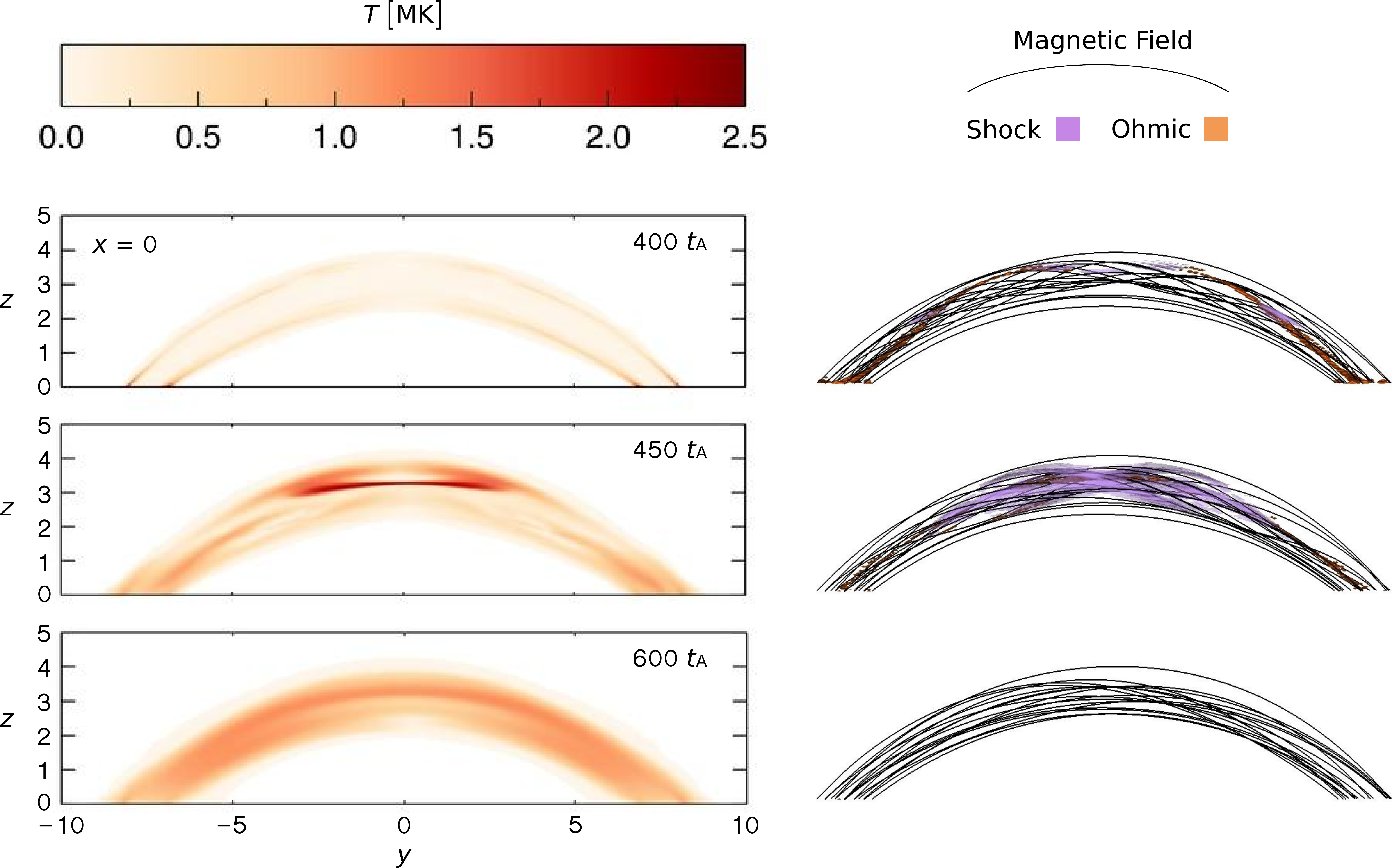}
  \caption{\small{\textbf{Loop C}, the temperature over a loop cross section (left column) taken at three times, instability onset (top), during the instability (middle) and the relaxed state (bottom). The configuration of the field lines, alongwith the pattern of Ohmic (orange) and shock (purple) heating (right column), are shown for the same times.}}
  \label{fig_temp_3dfield_c}
\end{figure}

\begin{figure}
  \center
  \includegraphics[width=0.75\textwidth,clip=]{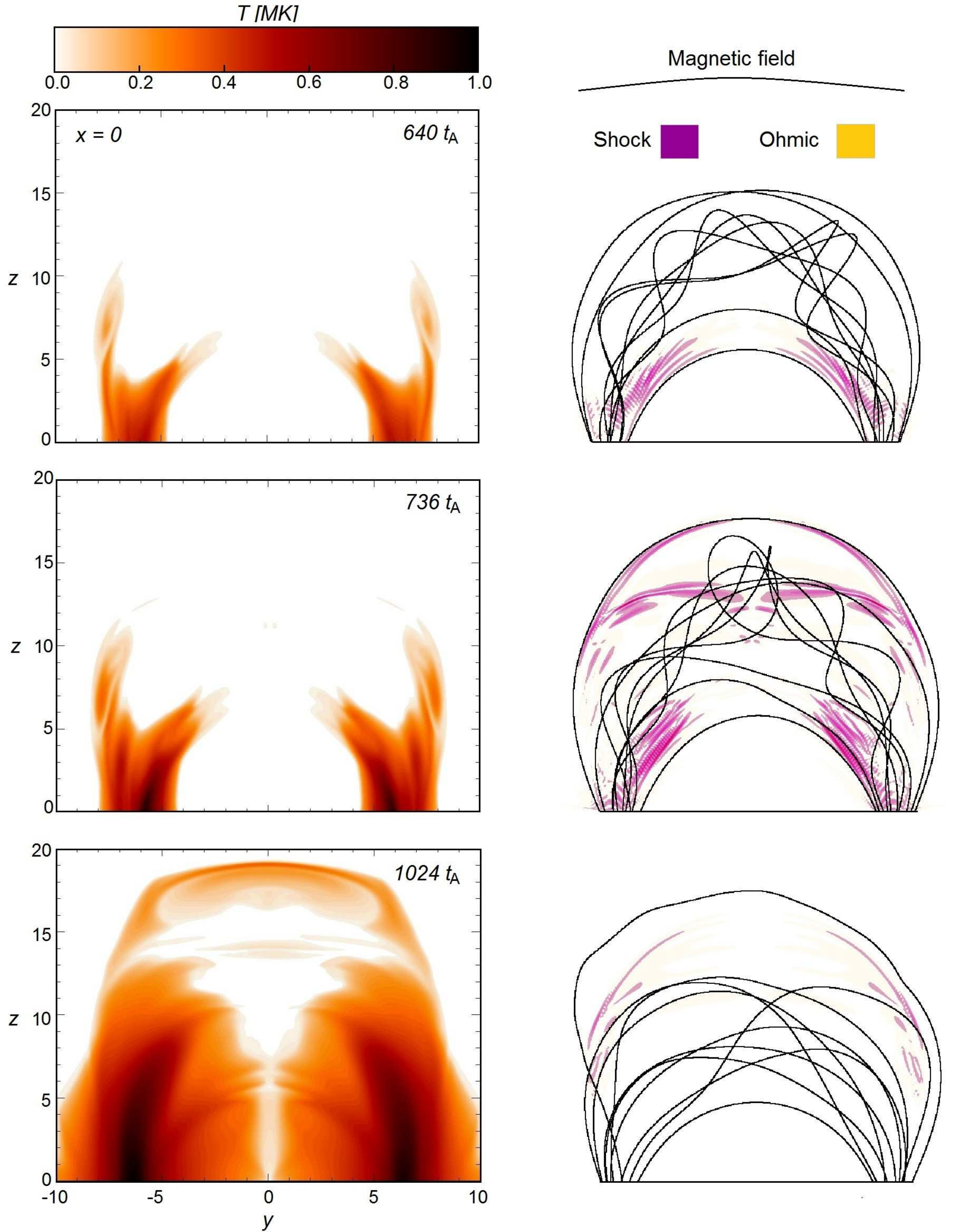}
  \caption{\small{\textbf{Loop D}, the format of this figure follows that used for Figure \ref{fig_temp_3dfield_c}. At $t\,{=}\,600\,t_{\mathrm{A}}$, the temperature plot is saturated: near the footpoints, the highest temperatures are $6\,\mathrm{MK}$.}}
  \label{fig_temp_3dfield_d}
\end{figure}

However, the magnetic fieldline plots in Figures \ref{fig_temp_3dfield_c} and \ref{fig_temp_3dfield_d} (right columns) do agree with the form expected for a kink instability: initially, the fieldlines are tightly twisted around the loop axis, and then, as the instability proceeds, the fieldlines untwist. \mbox{Loop C} undergoes an internal kink instability, since the apex height ($z\,{\approx}\,3.8$) remains almost constant throughout the simulation, and the heating (a mixture of Ohmic and shock) is concentrated around the apex during the unstable phase. As the loop relaxes, conduction combined with continued heating results in a near-uniform temperature of around $1.5\,\mathrm{MK}$. The temperature plots for loop D do suggest an increase in apex height. On the other hand, the corresponding fieldline plots indicate that the apex rise is temporary and is no longer evident once the loop has relaxed. Figure \ref{fig_temp_3dfield_d} also shows that the (mostly shock) heating is concentrated along the legs of the loop, and, for the scaling used here ($B_0\,{=}\,44.7\,\mathrm{G}$), results in sub-MK temperatures. There is some Ohmic heating at instability onset ($t\,{=}\,600\,t_{\mathrm{A}}$), but it is confined to the footpoints.

Comparison of all four models shows that the ratio of viscous to Ohmic dissipation changes from $\sim 1$ for a non-converging fluxtube \cite[see also][]{hooe09} to as much as $\sim 10$ in 
the strongly converging loop in model D. While the contribution of the Ohmic and viscous heating in models A and B is nearly uniform along the fluxtubes, in curved fluxtubes in models C and D the Ohmic heating is prevalent very close to footpoints (where the current density is high), while in the middle of the loops, where velocitites and velocity gradients are high, the magnetic energy dissipates through the ``viscosity'' channel. Hence, the enhanced viscous dissipation in converging fluxtubes is most likely caused by shocks formed due to low magnetic field and, hence, low Alfven velocities near the fluxtube centres.

 Strong shock dissipation in model D can explain the delay of the kink instability. Since the shock viscosity effectively acts as an additional Ohmic dissipation (see Section~\ref{sec_shock_res}), the rates of helicity injection and magnetic energy accumulation (due to the footpoint rotation) are lower. However, as far as the reconnection stage is concerned, slower energy release is due to the field convergence: lower magnetic field near the looptop and, hence, lower current densities, reduce the Ohmic heating rate, which is proportional to $J^2$.

\subsection{Stratified atmosphere}
\label{sec_stratified_atmosphere}

\begin{figure}[h!]
  \center
  \includegraphics[width=0.9\textwidth,clip=]{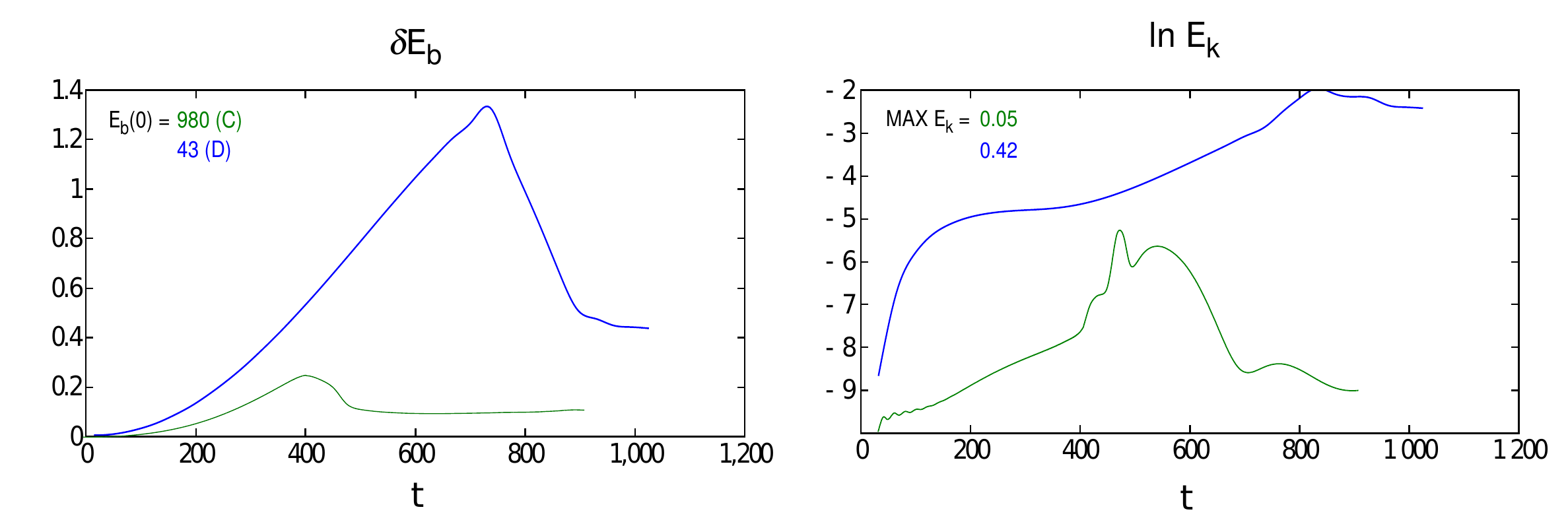}
  \caption{\small{The change in magnetic energy (left) and the natural logarithm of the kinetic energy (right) for loop C (green) and for loop D (blue). Resistive MHD was switched on at $t\,{=}\,400\,t_{\mathrm{A}}$ for loop C and $t\,{=}\,600\,t_{\mathrm{A}}$ for loop D, \ie, just before instability. Energies are volume intergrated.}}
  \label{fig_en_db_curved}
\end{figure}
\begin{figure}[h!]
  \center
   \includegraphics[width=0.55\textwidth,clip=]{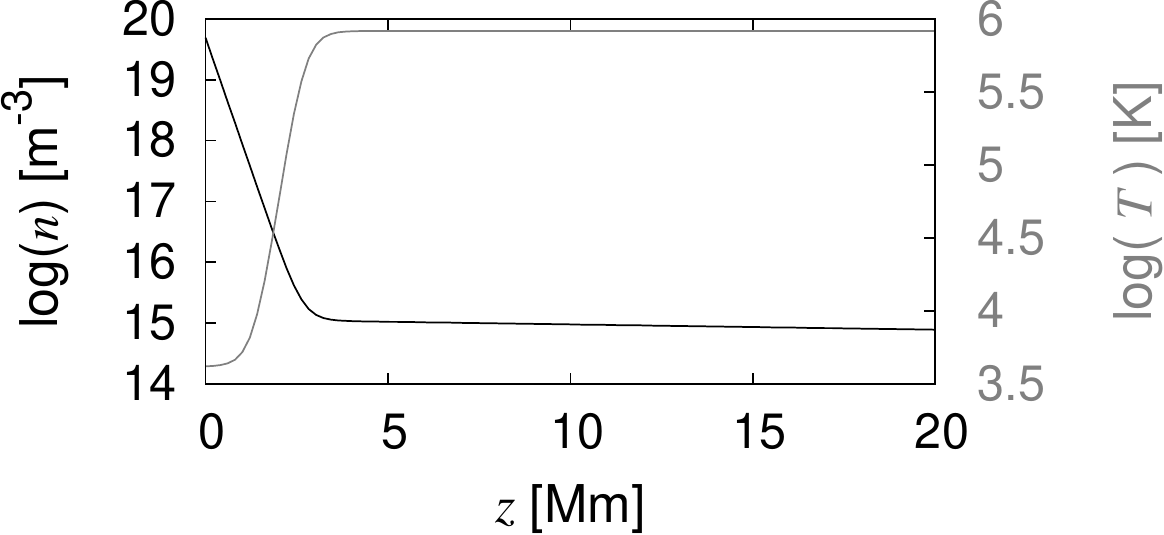}
  \caption{\small{The initial density (black) and temperature (grey) with height ($z$) for loop D*.}}
  \label{fig_atmos}
\end{figure}
\begin{figure}[h!]
  \center
  \includegraphics[width=0.9\textwidth,clip=]{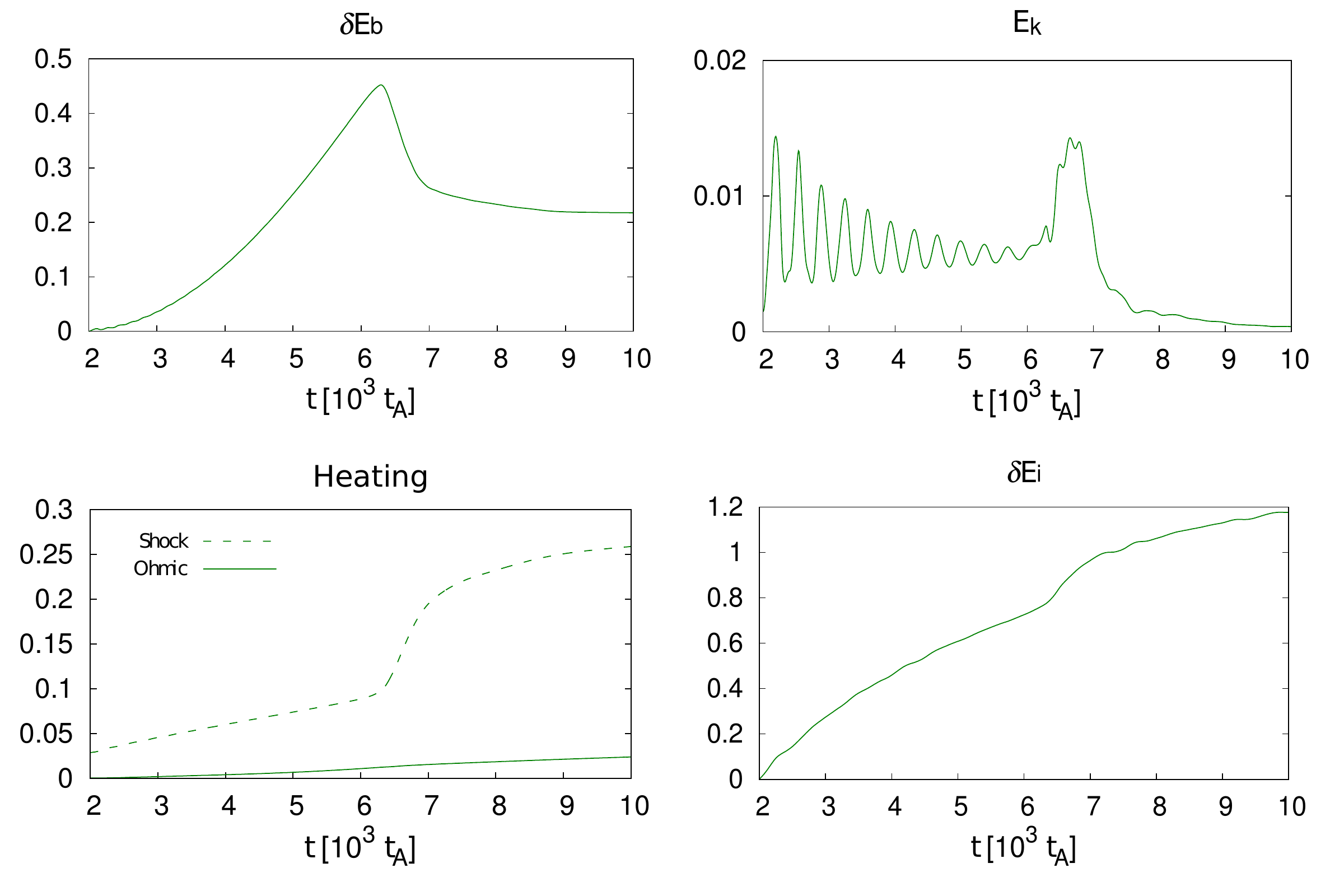}
  \caption{\small{The change in magnetic energy (top left), the kinetic energy (top right), the cumulative Ohmic and shock heating (bottom left), and the change in internal energy (bottom right) for loop D*gravitation. Energies are volume intergrated.}}
  \label{fig_en_curved_ds}
\end{figure}
\begin{figure}[h!] 
  \center 
  \includegraphics[width=0.55\textwidth,clip=]{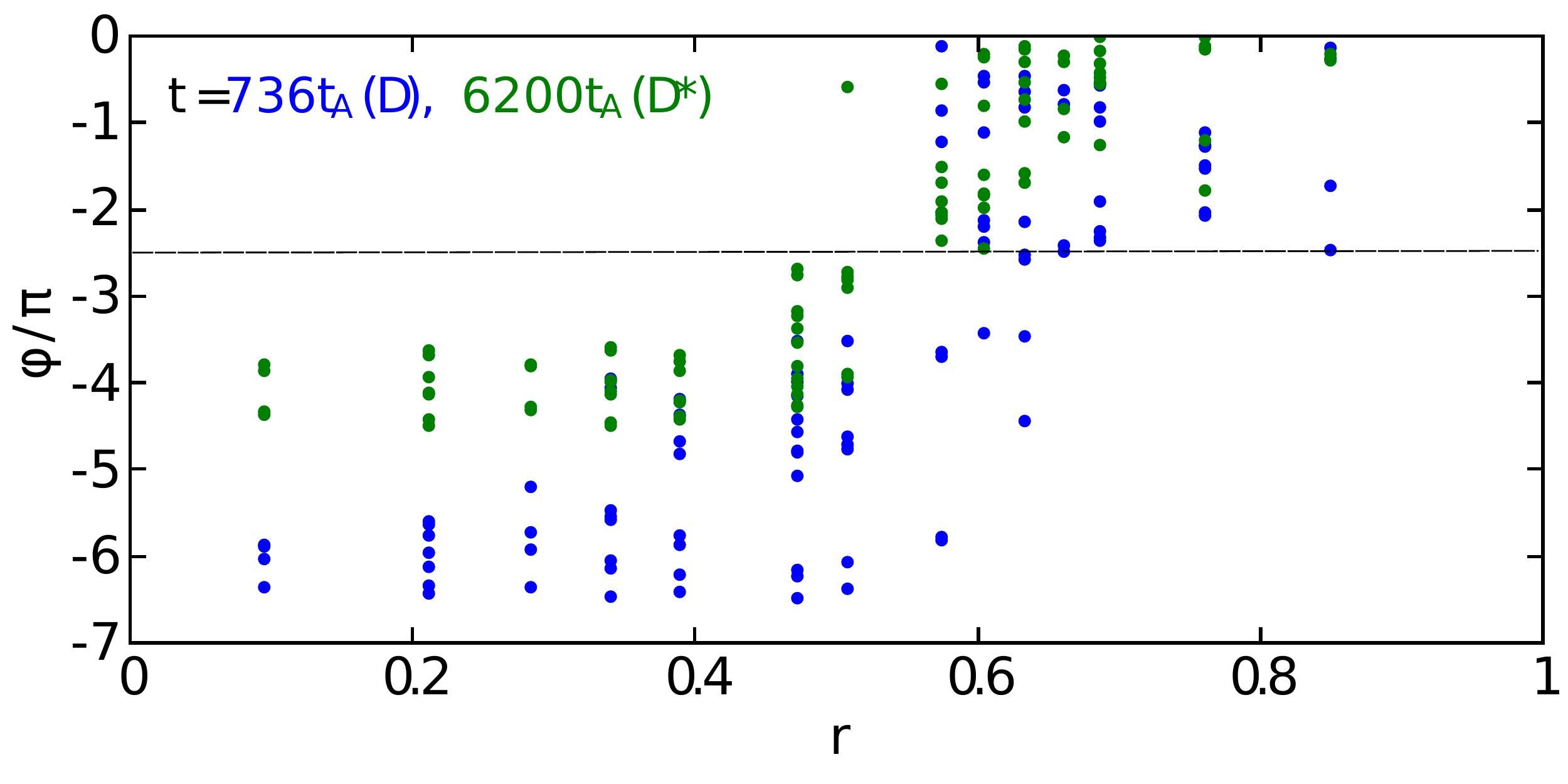}
  \caption{\small{The magnetic twist in units of $\pi$ as a function of radial distance from footpoint centre for loops, D (blue) and D* (green). The format of this figure follows that used for Figure \ref{fig_twist}.}}
  \label{fig_twist2}
\end{figure}

The loops mentioned so far all have a density and temperature that are initially uniform, where $n\,{=}\,1.2\times10^{15}\,\mathrm{m}^{-3}$ and $T\,{=}\,4\times10^3\,\mathrm{K}$. In order to investigate the effect of a stratified atmosphere, we also have loop D*, identical to loop D, except that it has a non-uniform atmosphere based on the work by \citet{gore13}: 
\begin{eqnarray}  
  \rho(z) & = & \rho_1\,\exp\Bigg(\frac{-(z_{\mathrm{sh}}+z)}{z_1}\Bigg) \,\,+\,\, \rho_2\,\exp\Bigg(\frac{-(z_{\mathrm{sh}}+z)}{z_2}\Bigg)\,,
  \label{eqn_atmos}  
\end{eqnarray}
where $\rho_1\,{=}\,3.34\times10^{-5}\,\mathrm{kg}\,\mathrm{m}^{-3}$ is the photospheric density, $\rho_2\,{=}\,2\times10^{-12}\,\mathrm{kg}\,\mathrm{m}^{-3}$ is the mean density in the corona, $z_1\,{=}\,0.25\,\mathrm{Mm}$ is the density scale height between the photosphere and the transition region, $z_2\,{=}\,50\,\mathrm{Mm}$ is the coronal scale height and $z_{\mathrm{sh}}\,{=}\,1.5\,\mathrm{Mm}$ simply allows the density profile to be shifted horizontally.

The initial atmosphere is shown in Figure \ref{fig_atmos} --- the temperature profile follows from hydrostatic balance, and so a gravitational term ($\,\rho\,g(z)\,$) is added to the force equation (\ref{eqn_lare_mhd_force}). Between $z\,{=}\,0$ and $z\,{=}\,4$ the temperature increases by more than two orders of magnitude (from around $4000\,\mathrm{K}$ to $0.83\,\mathrm{MK}$), while the particle number density drops from $10^{20}$ to $10^{15}\,\mathrm{m}^{-3}$. This region represents the chromosphere and transition region. The rest of the domain ($z\,{>}\,4$) represents the corona, where the temperature and density are more or less constant. It is appropriate to add an atmosphere to loop D, since the increase in thermal pressure towards the footpoints is matched by the hundredfold increase in magnetic pressure.

Increasing the density reduces the Alfv\'en speed, which has the unwelcome consequence of violating the constraint for direct current heating. At the footpoints, the driving speed (Figure \ref{fig_drv_spd}, left) is now twice the Alfv\'en speed, and so the perturbations generated by the driving occur too frequently for the loop to have time to settle into an equilibrium. This issue is easily rectified if the driving factor, $\omega_0$, is reduced by an order of magnitude. Hence, $v_{\mathrm{A}}(z\,{=}\,0)$ is now five times the driving speed, which compares well with observed values for active region field strengths ($0.1\,\mathrm{T}$), photospheric flow speeds ($1\,\mathrm{km}\,\mathrm{s}^{-1}$) and densities ($4\times 10^{-4}\,\mathrm{kg}\,\mathrm{m}^{-3}$).

The stratified atmosphere used for loop D* is not in equilibrium in the presence of thermal conduction; however, such an equilibirum can be achieved by allowing the simulation to evolve without driving for $2000\,t_{\mathrm{A}}$. Only then do we begin the driving phase with $t_{\mathrm{tw}}\,{=}\,4300\,t_{\mathrm{A}}$ set such that the driving ramps down at $6300\,t_{\mathrm{A}}$. Note, no restart is required after the driving phase, since in order for the atmosphere to acquire its initial equilibirum, the simulation must have resistive MHD and conduction switched on from the start.

The following energy plots begin from when the driving is started.

Figure \ref{fig_en_curved_ds} (top left) shows that the driving takes $4300\,t_{\mathrm{A}}$ to induce an instability. Comparison with loop D (Figure \ref{fig_en_db_curved}), reveals that the presence of an atmosphere requires less buildup in magnetic energy for instability onset, and so the resulting energy release is half that seen for loop D. This also means that adding a stratified atmosphere reduces the twist required for the instability onset: just before the kink instability loop D* has a lower twist for $r \leq 0.5$ compared to loop D.
The kinetic energy plot for loop D* exhibits the expected peak at the time the instability occurs, but there is also a wave pattern that, during the driving phase, decays in amplitude. Figure \ref{fig_en_curved_ds} (bottom left) shows that shock heating continues to dominate over Ohmic heating. Internal energy undergoes a continual increase, the gradient of which steepens as the loop goes unstable. The heating before the instability is possibly connected to the decaying wave form seen in the kinetic energy plot. Although disguised by the resolution of the time axis, the relaxation phase is even more extended than it is for loop D. The relaxation time is roughly $1000\,t_{\mathrm{A}}$, which equates to a dimensionalised time of $355\,\mathrm{s}$ if one uses the scalings given in Section \ref{sec_num_code}.

\section{Summary and conclusions}
\label{sec_summary_conclusions}

We have shown that the application of vortical driving to the footpoints of an initially potential field does lead to a kink instability, for straight and curved loops, with or without a stratified atmosphere. The rate at which the driving adds free magnetic energy to the loop is certainly enhanced when the field is made to converge at the foopoints. Thus, although converging loop B requires more energy to become unstable compared to its cylindrical counterpart (A), the buildup of energy is sufficiently fast that, for the same driving profile, loop B achieves instability around $100\,t_{\mathrm{A}}$ before loop A does. When comparing the curved loops (C and D), we again see that the strongly converging loop D has proportionally more free energy when instability occurs. Although, unlike the pair of loops A and B, the strongly converging loop D requires higher twist to become unstable compared to the weekly converging loop C. However, this case might be not representative, as the substantial viscous heating could affect the loop evolution prior to the instability.

The strongly converging loop D* (which is similar to the loop D, but embedded into the gravitationally stratified atmosphere) reaches the kink instability with a total twist nearly equal that in the loop C. Furthermore, taking into account the difference in twisting speed, it appears that they require the same amount of foot-point rotation to get the kink instability.
Hence, one can conclude (by comparing models A and B) that the field convergence reduces the stability of a twisted loop in cylindrically symmetric magnetic fluxtubes. However, the convergence does not seem to influence the stability of the fluxtubes with large-scale curvature. These conclusions with regard to the loop stability should be used with caution, as the stability can be influenced by many effects, which can't be examined using only five test models.

The large-scale curvature affects the geometry of magnetic reconnection. The current distribution loses its cylindrical symmetry, with strong currents formed in a thin shield above the loop top. Although, physically it is represented by two effects -- twist reduction and fluxtube expansion (due to the reconnection with ambient field) -- the reconnection is more localised around the central part of the loop, leading to topological evolution descibed by 
\citep{gore14}.

As regards heating, loops A and C show equal amounts of Ohmic and shock heating, whereas loops B, D and D* are mostly heated through shocks. The shock heating term \mbox{(Section \ref{sec_shock_res})} in the energy equation becomes more significant when higher energies are required to drive the loop unstable. The dominance of shock heating is also true when there is a stratified atmosphere. The photospheric density for D* is some five orders of magnitude higher than the initially uniform density used for loop D. For this reason, the driving speed is reduced by a factor of ten, so that it is five times lower than the Alfv\'en speed at $z\,{=}\,0$, consistent with observed values. These two speeds are sufficiently close for wave phenomena to be visible in the kinetic energy profile. The decaying nature of these waves suggests that some heating of the plasma is occurring during the driving phase (Figure \ref{fig_en_curved_ds}, top right), which perhaps destablises the loop.

The important methodological implication from our result is that the shock viscosity needs to be taken into account as an additional Ohmic dissipation. This issue is particularly important for numerical experiments with localised resistivity effects, as shocks would affect effective spatial resistivity distribution.

\begin{acknowledgements}
This work is funded by Science and Technology Facilities Council (UK). The simulations were run on the UK MHD Consortium cluster based in St Andrews and on the COSMA Data Centric system at Durham University. The latter is operated by the Institute for Computational Cosmology on behalf of the STFC DiRAC HPC Facility (www.dirac.ac.uk). This equipment was funded by a BIS National E-infrastructure capital grant ST/K00042X/1, DiRAC Operations grant ST/K003267/1 and Durham University. DiRAC is part of the National E-Infrastructure. 
\end{acknowledgements}

\end{article} 


\begin{thebibliography}{00}
\bibitem[\protect\citeauthoryear{Arber \etal}{2001}]{arbe01} Arber, T.G., Longbottom, A.W., Gerrard, C.L., Milne, A.M.: 2001, {\it J. Comp. Phys.} {\bf 171}, 151.
\bibitem[\protect\citeauthoryear{Bareford \etal}{2010}]{bare10} Bareford, M.R., Browning, P.K., Van der Linden, R.A.M.: 2010, {\it Astron. Astrophys.} {\bf 521}, 70.
\bibitem[\protect\citeauthoryear{Bareford \etal}{2011}]{bare11} Bareford, M.R., Browning, P.K., Van der Linden, R.A.M.: 2011, {\it Solar Phys.} {\bf 273}, 93.
\bibitem[\protect\citeauthoryear{Bareford \etal}{2013}]{bare13} Bareford, M.R., Hood, A.W., Browning, P.K.: 2013, {\it Astron. Astrophys.} {\bf 550}, 40.
\bibitem[\protect\citeauthoryear{Bareford and Hood}{2015}]{bare14} Bareford, M.R., Hood, A.W.: 2015, {\it Phil. Trans. Roy. Soc. A.} {\bf 373}, 20140266.
\bibitem[\protect\citeauthoryear{Botha \etal}{2011}]{bote11} Botha, G.J.J., Arber, T.D., Hood, A.W.: 2011, {\it Astron. Astrophys.} {\bf 525}, A96.
\bibitem[\protect\citeauthoryear{Braginskii}{1965}]{brag65} Braginskii, S.I.: 1965, {\it Rev. Plasma Phys.} {\bf 1}, 205.
\bibitem[\protect\citeauthoryear{Brio and Wu}{1998}]{brio98} Brio, M., Wu, C.C.: 1998, {\it J. Comp. Phys.} {\bf 75}, 400.
\bibitem[\protect\citeauthoryear{Browning and Van der Linden}{2003}]{brva03} Browning, P.K., van der Linden, R.A.M.: 2003, {\it Astron. Astrophys.} {\bf 400}, 355.
\bibitem[\protect\citeauthoryear{Browning \etal}{2008}]{broe08} Browning, P.K., Gerrard, C., Hood, A.W., Kevis, R., van der Linden, R.A.M.: 2008, {\it Astron. Astrophys.} {\bf 485}, 837.
\bibitem[\protect\citeauthoryear{Cirtain \etal}{2013}]{cire13} Cirtain, J.W., Golub, L., Winebarger, A.R., de Pontieu, B., Kobayashi, K., Moore, R.L., Walsh, R.W., Korreck, K.E., Weber, M., McCauley, P., Title, A., Kuzin, S., Deforest, C.E.: 2013, {\it Nature} {\bf 493}, 501.
\bibitem[\protect\citeauthoryear{Gordovskyy \etal}{2013}]{gore13} Gordovskyy, M., Browning, P.K., Kontar, E.P., Bian, N.H.: 2013, {\it Solar Phys.} {\bf 284}, 489.
\bibitem[\protect\citeauthoryear{Gordovskyy \etal}{2014}]{gore14} Gordovskyy, M., Browning, P.K., Kontar, E.P., Bian, N.H.: 2014, {\it Astron. Astrophys.} {\bf 561}, 72.
\bibitem[\protect\citeauthoryear{Gordovskyy and Browning}{2011}]{gobr11} Gordovskyy, M., Browning, P.K.: 2011, {\it Astrophys. J.} {\bf 729}, 101.
\bibitem[\protect\citeauthoryear{Gordovskyy and Browning}{2012}]{gobr12} Gordovskyy, M., Browning, P.K.: 2012, {\it Solar Phys.} {\bf 277}, 299.
\bibitem[\protect\citeauthoryear{Hood and Priest}{1979}]{hopr79} Hood, A.W., Priest, E.R.: 1979, {\it Solar Phys.} {\bf 64}, 303.
\bibitem[\protect\citeauthoryear{Hood}{1992}]{hood92} Hood, A.W.: 1992, {\it Plasma Phys. Contr. Fusion} {\bf 34}, 411.
\bibitem[\protect\citeauthoryear{Hood \etal}{2009}]{hooe09} Hood, A.W., Browning, P.K., Van der Linden, R.A.M.: 2009, {\it Astron. Astrophys.} {\bf 506}, 913.
\bibitem[\protect\citeauthoryear{Klimchuk}{2000}]{klim00} Klimchuk, J.A.: 2000, {\it Solar.Phys.} {\bf 193}, 53.
\bibitem[\protect\citeauthoryear{Kumar and Cho}{2014}]{kuch14} Kumar, P., Cho, K.S.: 2014, {\it Astron. Astrophys.} {\bf 572}, A83.
\bibitem[\protect\citeauthoryear{Kuridze \etal}{2013}]{kure13} Kuridze, D., Mathioudakis, M., Kowalski, A.F., Keys, P.H., Jess, D.B., Balasubramaniam, K.S., Keenan, F.P.: 2013, {\it Astron. Astrophys.} {\bf 552}, A55.
\bibitem[\protect\citeauthoryear{Mellor \etal}{2005}]{mele05} Mellor, C, Gerrard, C.L., Galsgaard, K., Hood, A.W., Priest, E.R.: 2005, {\it Solar Phys.} {\bf 227}, 39.
\bibitem[\protect\citeauthoryear{Parker}{1988}]{park88} Parker, E.N.: 1988, {\it Astrophys. J.} {\bf 330}, 474.
\bibitem[\protect\citeauthoryear{Peter and Bingert}{2012}]{pebi12} Peter, H., Bingert, S.: 2012, {\it Astron. Astrophys.} {\bf 548}, 1.
\bibitem[\protect\citeauthoryear{Parnell and De Moortel}{2012}]{pade12} Parnell, C.E., De Moortel, I.: 2012, {\it Phil. Tran. Roy. Soc. A} {\bf 370}, 3217
\bibitem[\protect\citeauthoryear{Reale}{2014}]{real14} Reale, F.: 2014, {\it Liv. Rev. Solar Phys.} {\bf 11}.
\bibitem[\protect\citeauthoryear{Srivastava \etal}{2010}]{srie10} Srivastava, A.K., Zaqarashvili, T.V., Kumar, P., Khodachenko, M.L.: 2010, {\it Astrophys. J.} {\bf 715}, 292.
\bibitem[\protect\citeauthoryear{Van Leer}{1997}]{vanl97} Van Leer, B.: 1997, {\it J. Comput. Phys.} {\bf 135}, 229.
\bibitem[\protect\citeauthoryear{Wilkins}{1980}]{wilk80} Wilkins, M.L.: 1980, {\it J. Comput. Phys.} {\bf 36}, 281.
\end{thebibliography}
\end{document}